\documentclass{aa}   
\usepackage{CJKutf8}
\usepackage{graphicx}
\usepackage[breaklinks,colorlinks,citecolor=blue,linkcolor=blue]{hyperref}
\usepackage{txfonts}
\usepackage{lscape}
\usepackage{epstopdf}
\usepackage{CJK}
\usepackage{float}
\usepackage[absolute]{textpos}
\usepackage{amsmath}
\usepackage{mathrsfs}

\usepackage{arydshln}

\begin{document} 
   \title{PMO Polaris CO survey. II. Where is the dust?}
   \author{Xunchuan Liu \begin{CJK}{UTF8}{gbsn}(刘训川)\end{CJK}
          \inst{1}\thanks{liuxunchuan001@gmail.com}
          \and 
          Bing-Gang Ju\inst{2}
          \and 
          Fujun Du\inst{2}
          \and 
          Paul F. Goldsmith\inst{3}
          \and
          Tianwei Zhang\inst{4}
          \and 
          Lixia Yuan\inst{2}
          \and
                    Lianghao Lin\inst{5}
          \and
          Zhihong He\inst{6}
          \and 
          Chao Zhang\inst{7}
          \and 
          Ping Yan\inst{2}
          \and 
          Shengyu Jin\inst{2}
          \and 
          Yongxing Zhang\inst{2}
          \and 
          Dengrong Lu\inst{2}
          } 

   \institute{
Leiden Observatory, Leiden University, P.O. Box 9513, 2300RA Leiden, The Netherlands \and
Purple Mountain Observatory, Chinese Academy of Sciences, Nanjing 210023 \and
Jet Propulsion Laboratory, California Institute of Technology, 4800 Oak Grove Drive, Pasadena, CA 91109, USA
              \and 
Research Center for Computational Earth and Space Science, Zhejiang Laboratory, Hangzhou 311100, China
\and
Max-Planck-Institut f\"ur Radioastronomie, Auf dem Hügel 69, 53121 Bonn, Germany
\and 
School of Physics and Astronomy, China West Normal University, 
No. 1 Shida Road, Nanchong 637002, China
\and 
Institute of Astronomy and Astrophysics, School of Mathematics and Physics, Anqing Normal University, Anqing, China
              }
              
\date{Received xxx xx}

\abstract{
Dust plays critical chemical and dynamical roles in the interstellar medium (ISM), but its specific association with molecular and atomic gas remains difficult to isolate. Combining the PMO Polaris CO Survey (PPCOS), EBHIS \ion{H}{I} data, and \textit{Planck} dust maps, this study investigates dust distributions across multiple gas components in the Polaris Flare. We employ multi-technique linear decomposition---including full-spectrum fitting and a regularization approach---to reconstruct the dust distribution from multi-component gas emissions.
This framework quantifies dust contributions from CO-associated, \ion{H}{I}-associated, and CO-dark molecular gas phases. CO-associated dust accounts for 20--40\% of the total dust mass, whereas dust in the broad \ion{H}{I} (warm neutral medium, WNM) component is negligible. Instead, \ion{H}{I}-associated dust concentrates primarily within the narrow cold neutral medium (CNM) and a distinct, ultra-narrow component with a velocity width comparable to the \ion{H}{I} spectral resolution.
Residual dust at atomic-to-molecular (\ion{H}{I}--CO) interfaces contributes 4--10\% to the global dust mass, but exceeds 25\% at molecular cloud boundaries, confirming a substantial presence of CO-dark molecular gas. Furthermore, the velocity fields of dust-associated \ion{H}{I} closely match those of CO, indicating active dynamical coupling between CO-emitting gas and the surrounding CNM.
Guided by these results, we present a stepwise schematic cartoon illustrating the coupling between multi-phase gas structures, molecular formation, and dust growth.
}


 \maketitle
 

\section{Introduction} \label{sec_intro}
Dust grains are ubiquitous in the interstellar medium (ISM), regulating heating, cooling, extinction, cloud formation, and chemical processes such as molecular hydrogen and complex organic molecule synthesis \citep[e.g.,][]{1971ApJ...163..155H,1990ARA&A..28...37M,2003ARA&A..41..241D,2009ARA&A..47..427H,2020ARA&A..58..529S}. Dust emission broadly traces atomic and molecular gas distributions \citep[e.g.,][]{1997ApJ...491..615G,2007A&A...465..839P,2015A&A...582A..31P} and can reveal material otherwise difficult to detect, including CO-dark molecular gas \citep{2012A&A...543A.103P,2014MNRAS.441.1628S}. Establishing dust distributions among these phases, as well as within residual components, is therefore crucial to understanding the physical state, structure, and evolution of the ISM.

However, quantitatively separating dust associated with distinct gas phases remains challenging because observed dust emission represents a line-of-sight superposition of molecular, atomic, and transitional components \citep{2007A&A...465..839P,2020A&A...639A..26K}. Molecular gas can be effectively probed using CO and its isotopologues, which trace different density regimes and constrain the associated dust emission \citep[e.g.,][]{2010ApJ...721..686P}. Concurrently, atomic gas consists of the warm neutral medium (WNM) and the cold neutral medium (CNM), both of which may exhibit distinct dust properties and spatial distributions \citep[e.g.,][]{2018A&A...619A..58K}.
Consequently, large-area CO, \ion{H}{I}, and dust observations of structurally simple regions are essential to quantitatively and statistically decouple dust contributions across disparate gas phases.

While all-sky \ion{H}{I} and dust surveys provide extensive Galactic coverage \citep{2014A&A...571A..11P,2016A&A...585A..41W,2025ApJS..280...15W}, allocating dust to atomic or molecular components remains difficult without high-resolution CO data, particularly at high latitudes and in low-density regimes. Historically, most high-resolution CO surveys focused on the Galactic plane \citep[e.g.,][]{2006ApJS..163..145J,2019ApJS..240....9S}, where crowded, complex environments hinder dust phase separation. Conversely, the CfA CO survey \citep{2001ApJ...547..792D} lacks the spatial resolution and Nyquist sampling required for detailed dust decomposition at high latitudes \citep{1990ApJ...353L..49H}. Absent adequate CO data, dust decomposition typically overestimates the atomic gas contribution, leaving the dust associated with CO-dark molecular gas unconstrained \citep[e.g.,][]{2012A&A...543A.103P}.

The PMO Polaris CO Survey (PPCOS; see \citealt{liu2026pmopolarissurveyi}, Paper~I) helps resolve this limitation by delivering the first large-area (100~deg$^2$), high-latitude ($|b|>20^\circ$) CO(1--0) map with arcminute resolution and proper Nyquist sampling. As a relatively isolated, cold, and evolutionarily young cloud devoid of star formation \citep{1998A&A...333..709L,1999A&A...347..640B,2010A&A...518L.102A}, the Polaris Flare serves as an ideal laboratory to study the atomic--molecular interface. Integrating PPCOS with EBHIS \ion{H}{I} data and \textit{Planck} dust maps enables a robust, quantitative linear decomposition. This combination prevents the overestimation of dust associated with \ion{H}{I}, allows precise determination of CO-associated dust, and provides a unique baseline to isolate and characterize CO-dark molecular gas.
In particular, comparing spectrally resolved, dust-related \ion{H}{I} with CO emission enables us to explore the coherence of multi-phase structures. This comparison allows us to assess how dynamical processes, such as large-scale velocity gradients and turbulence, shape the observed gas and dust distributions.

The paper is structured as follows. Section~\ref{sec_data} describes the CO, \ion{H}{I}, and dust datasets used in this work. Section~\ref{over} presents the objective dust map together with the fitted images of CO and the narrow and broad \ion{H}{I} components. Section~\ref{sec_method} outlines the linear fitting method adopted in this study. The results from fitting the CO and \ion{H}{I} moment maps are given in Sect.~\ref{sec_basicdecomp}, while Sect.~\ref{sec_fullspecfit} reports the results of full-spectrum fitting, which employs \ion{H}{I} channel maps as templates and applies regularized linear fitting to identify the channels contributing to the dust emission. Section~\ref{sec_reg_phy} examines how these channels account for the dust emission and reproduce CO-like dynamics. Finally, Sects.~\ref{sec_diss} and \ref{sec_summary} provide a brief discussion and summary.


\section{Data}\label{sec_data}
This study utilizes CO ($J=1\!-\!0$) emission data from the PMO Polaris CO survey, dust maps derived from the \textit{Planck} continuum \citep{2014A&A...571A..11P}, and \ion{H}{I} 21~cm line data from the EBHIS survey \citep{2016A&A...585A..41W}.

\begin{figure}
    \centering
    \includegraphics[width=0.98\linewidth]{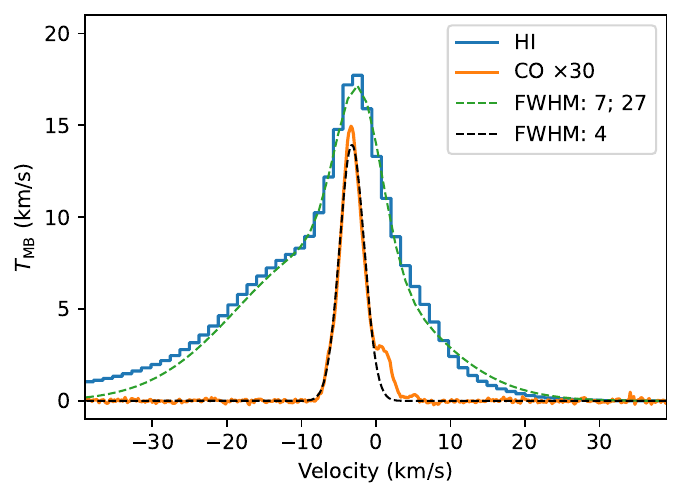}
    \caption{Spectra of CO (PPCOS; Liu et al. 2025) and 
\ion{H}{I} \citep[EBHIS;][]{2016A&A...585A..41W}, 
averaged over the PPCOS-mapped region of the Polaris Flare.
The dashed lines show Gaussian fits. 
To match the intensity of \ion{H}{I} (blue), the CO spectrum (orange)
has been multiplied by a factor of 30.
The FWHM line widths of the narrower and broader components of \ion{H}{I} 
are 7 km~s$^{-1}$ and 27 km~s$^{-1}$, respectively,
while the FWHM of CO is 3.9 km~s$^{-1}$.}
    \label{fig:meanspec_COHI}
\end{figure}

\begin{figure*}
    \centering
    \includegraphics[width=0.97\linewidth]{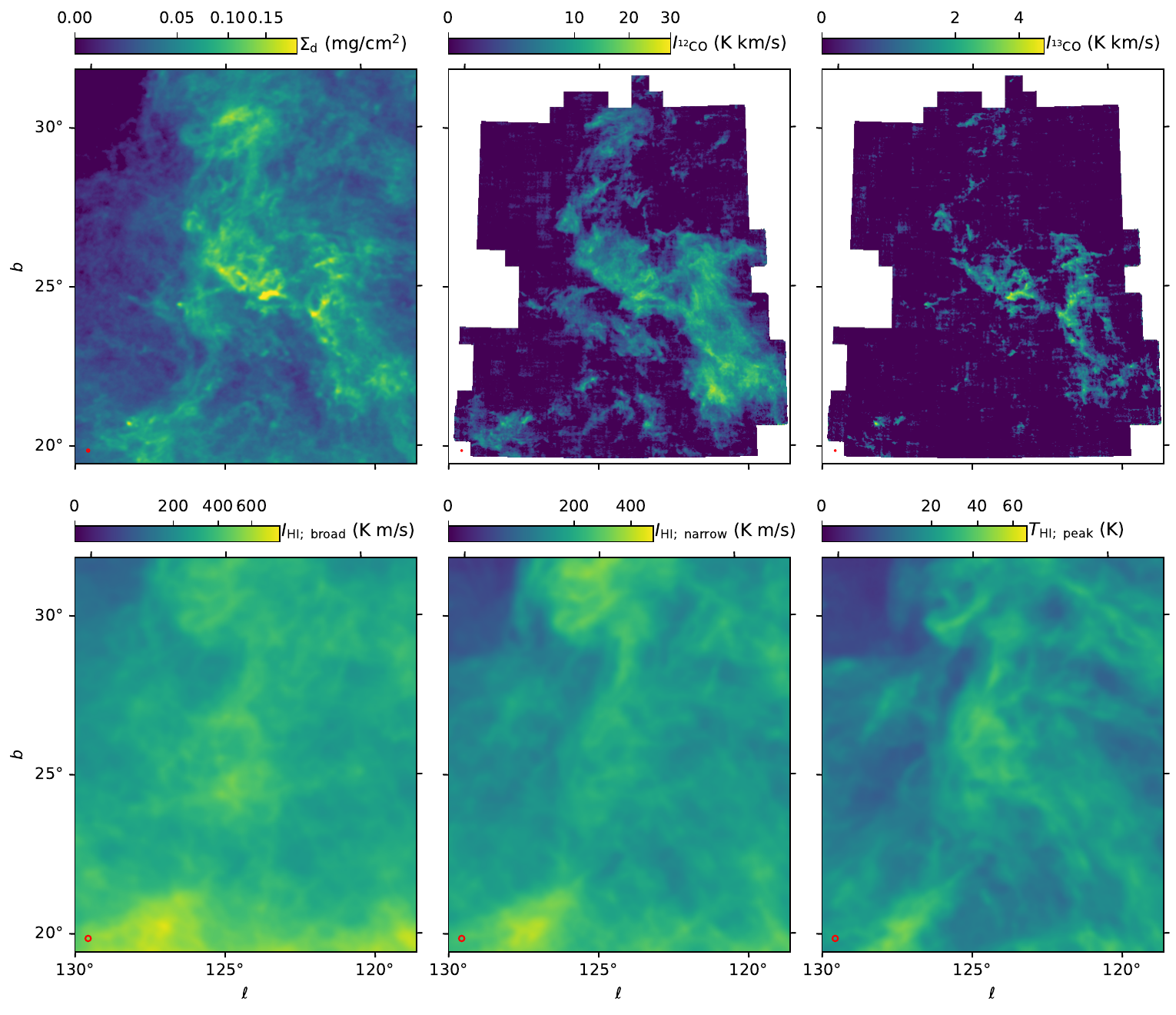}
    \caption{Maps of the surface density of Planck dust ($\Sigma_d$), 
    the integrated intensities of CO lines ($I_{\rm ^{12}CO}$ and $I_{\rm ^{13}CO}$), 
    and \ion{H}{I} lines. The \ion{H}{I} maps include the broad component $I_{\rm HI, broad}$ (from $-40$ to $20$ km s$^{-1}$), 
    the narrow component $I_{\rm HI, narrow}$ (from $-8$ to $5$ km s$^{-1}$), 
    and the intensity at $-2.5$ km s$^{-1}$, $T_{\rm HI, peak}$. 
    The colorbar labels indicate the names of the quantities shown in each map. 
    The CO maps have not yet been smoothed to match the spatial resolution of the dust map. 
    All integrated intensity maps are obtained by direct integration with no intensity threshold, 
    avoiding bias from missing weak, extended structures that would only become apparent after smoothing or in a statistical analysis.
    The effective spatial resolution of the maps is indicated by a red circle in the lower corner of each panel.
    \label{fig_overall}}
\end{figure*}

\begin{figure}[t]
    \centering
    \includegraphics[width=0.99\linewidth]{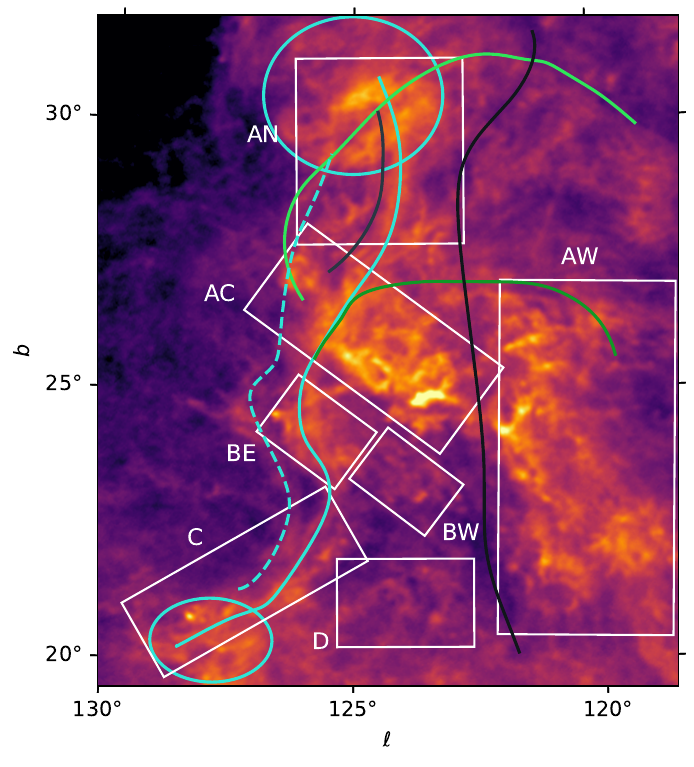}
    \caption{Features revealed from the dust map and spectral line maps. 
    Background is the dust map.
White boxes indicate the subcomplexes identified from the CO maps (Paper I). 
The cyan solid line traces the eastern edge, with two ellipses marking its two heads (clumps). 
The dashed cyan line highlights a fainter, slim structure (denoted as the faint edge) alongside the eastern edge (cyan solid line, Sect.~\ref{sect_imoverview}). 
Green lines mark the main C shape and the northern C shape visible on the CO-subtracted dust residual map (Sect.~\ref{sec_detrendcross}). 
The two black lines indicate the dust gaps.}
    \label{fig:marks}
\end{figure}

\subsection{PMO Polaris CO survey}
The PMO Polaris CO Survey (PPCOS; \citealt{liu2026pmopolarissurveyi}, Paper I) covers a $\sim$100~deg$^2$ area of the Polaris Flare in the $J=1\!-\!0$ transitions of $^{12}$CO, $^{13}$CO, and C$^{18}$O. Observations were conducted using the Delingha 13.7\,m telescope at the Purple Mountain Observatory, located at an elevation of $\sim3200$\,m (latitude $37^{\circ}22.4'\,\mathrm{N}$, longitude $97^{\circ}33.6'\,\mathrm{E}$), which ensures stable atmospheric conditions for millimeter-wavelength observations \citep{2016PASP..128j5003T}. Equipped with a $3 \times 3$ multibeam SIS receiver \citep{2012ITTST...2..593S} and high-resolution FFT spectrometers \citep{ZHENQIANG2021559}, the system simultaneously recorded all three CO transitions. The configured channel spacing of $\sim61$~kHz yields an exceptional velocity resolution of $\sim0.16$--$0.17$~km~s$^{-1}$.
The survey grid comprises 407 cells, nominally sized at $30'\times30'$ but expanded to $40'\times40'$ to mitigate noisy margins, mapping a net area of $\sim95.5$~deg$^2$ after accounting for spatial overlaps. Data acquisition utilized the position-switching On-The-Fly (OTF) mode to ensure strict Nyquist sampling across approximately 2000 total scans. This setup achieves a native angular resolution of $\sim50''$ for $^{12}$CO, with a slightly broader beam width for $^{13}$CO and C$^{18}$O. The resulting typical root-mean-square (rms) noise levels are 0.46~K per 0.16~km~s$^{-1}$ channel for $^{12}$CO and $\sim0.23$~K for both $^{13}$CO and C$^{18}$O.

PPCOS represents the first high-latitude ($b>20\degr$) sub-arcminute CO survey of this scale, whereas historical campaigns predominantly targeted the dense Galactic plane \citep{2006ApJS..163..145J,2015ApJ...812....6B,2017PASJ...69...78U,2019ApJS..240....9S}. For the present analysis, we smooth the merged data cube to an effective angular resolution of $\sim1.5\arcmin$ with a spatial gridding step of $0.5\arcmin$, which optimizes the structural resolution while establishing a more uniform noise distribution. Despite this smoothing, the PPCOS dataset retains an angular resolution substantially finer than the EBHIS \ion{H}{I} data and the \textit{Planck} thermal dust maps.
The CfA 1.2\,m Polaris survey \citep[][HT90]{1990ApJ...353L..49H,1993A&A...268..265H} mapped $^{12}$CO over $\sim50$~deg$^2$ toward the Polaris Flare, using an 8.7\arcmin\ half-power beam width sampled on a coarse 7.5\arcmin\ grid (roughly half-Nyquist sampling) with a spectral resolution of 0.65~km~s$^{-1}$. This coarse spatial sampling, low velocity resolution, and limited sensitivity render HT90 suboptimal for cross-correlation with \textit{Planck} dust maps. Furthermore, the sky coverage of HT90 is heavily biased toward regions with relatively strong CO emission, leaving it unable to span the extended dust structures. Additionally, the lack of $^{13}$CO emission limits its ability to trace the densest regions where dust emission is brightest. Such limitations prevent legacy datasets from decoupling dust-related gas emissions within clouds characterized by narrow line widths and weak molecular lines. Consequently, PPCOS is not merely a superior alternative to legacy surveys, but an indispensable asset for multi-phase ISM analysis. In contrast to historical constraints, the wide coverage, sub-arcminute native resolution, Nyquist-sampled OTF execution, and multi-isotopologue measurements of PPCOS permit a detailed characterization of molecular gas morphology and kinematics, establishing a precise baseline for cross-correlating multi-phase ISM components.

\subsection{EBHIS HI data}
Atomic hydrogen (\ion{H}{I}) 21~cm line data are obtained from the Effelsberg-Bonn \ion{H}{I} Survey \citep[EBHIS;][]{2016A&A...585A..41W,2016A&A...594A.116H}. EBHIS covers the northern sky ($\delta \gtrsim -5\degr$) at an angular resolution of 10.8$'$, a spectral channel spacing of 1.44~km~s$^{-1}$, and a typical root-mean-square (rms) noise level of $\sim$90~mK. This survey fully encompasses the Polaris Flare, capturing both diffuse, large-scale structures and localized emission features at high Galactic latitudes. Compared to the southern-sky GASS survey or the merged, all-sky HI4PI product—which are limited to a coarser 16.2$'$ resolution—the native EBHIS data preserve a higher spatial resolution. Although  HI4PI were widely
utilized for researches \citep[e.g.,][]{2024A&A...686A.305F,2025NatAs...9.1366L}, the resolution degradation non-negligible here for fine-scale decomposition.
While the 10.8$'$ resolution of EBHIS remains coarser than that of the PPCOS CO data and \textit{Planck} dust maps, it represents the highest-resolution atomic gas dataset available for this region. Unsteerable instruments, such as Arecibo \citep[zenith angle limit $\sim19.7^\circ$;][]{2005AJ....130.2598G} and FAST \citep[zenith angle limit $\sim40^\circ$;][]{2011IJMPD..20..989N}, cannot observe the high-declination Polaris Flare ($\delta \sim 75^\circ$). Consequently, EBHIS provides the optimal constraints on the atomic gas distribution for our analysis. 
To maximize the spatial information, we utilize the unsmoothed EBHIS \ion{H}{I} data throughout this work. The native data are reprojected from the public HEALPix cube\footnote{\url{https://cdsarc.cds.unistra.fr/ftp/J/A+A/585/A41/}} via the generalized bilinear interpolation procedure detailed in Appendix~\ref{sec_bilinear}.

\begin{figure*}[t]
    \centering
    \includegraphics[width=0.99\linewidth]{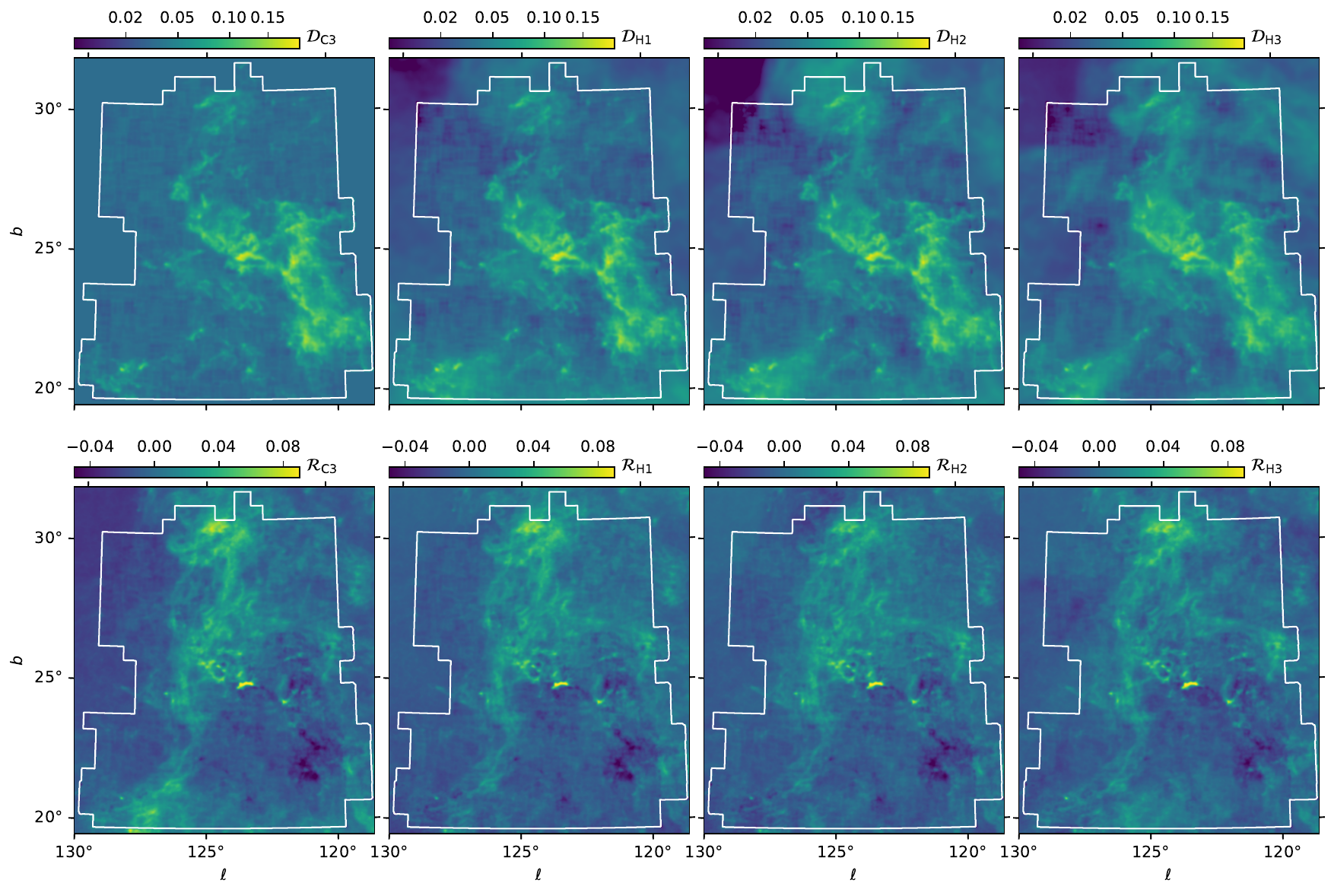}
    \caption{Fitted dust map (upper) and residual map (lower) for different fittings (Table~\ref{tab_fit}). 
The white line outlines the image region of the PMO Polaris CO survey (PPCOS).
    \label{fig_fitting}} 
\end{figure*}
\begin{table*}
    \caption{Linear decomposition results for different cases$^{(1)}$.\label{tab_fit}}
    \centering
    \begin{tabular}{lcccccccccc}
    \hline \hline 
   Case$^{(2)}$ & $I_{\rm ^{12}CO}$ & $I_{\rm ^{13}CO}$ & $I_{\rm C^{18}O}$ & $I_{\rm HI,\ broad}$ & $I_{\rm HI,\ narrow}$ & $T_{\rm HI,\ peak}$ & $I_c$ & 
    rms($\mathcal{D}_{\rm fit}$) & rms($\mathcal{D}_{\rm fit}-I_c$) & rms($\mathcal{R}$)$^{(3)}$ \\
    \hline
C1 & 0.0055 & -- & -- & -- & -- & -- & 0.0330 & 0.0479 & 0.0212 & 0.0167\\
C2 & 0.0041 & 0.017 & -- & -- & -- & -- & 0.0340 & 0.0481 & 0.0212 & 0.0162 \\
C3 & 0.0041 & 0.014 & 0.073 & -- & -- & -- & 0.0340 & 0.0481 & 0.0212 & 0.0161 \\
H1 & 0.0041 & 0.017 & 0.074 & 1.0E-04 & -- & -- & -0.0008 & 0.0488 & 0.0495 & 0.0139 \\
H1$^{\dag}$ & 0.0043 & 0.015 & 0.067 & 9.4E-05 & -- & -- & 0.0190 & 0.0483 & 0.0317 & 0.0156 \\
H2 & 0.0036 & 0.018 & 0.087 & -- & 2.0E-04 & -- & -0.0007 & 0.0491 & 0.0497 & 0.0130\\
H3 & 0.003 & 0.014 & 0.086 & -- & -- & 0.0016 & 0.0080 & 0.0491 & 0.0419 & 0.0130\\
H4 & 0.0036 & 0.018 & 0.088 & -1.2E-05 & 2.2E-04 & -- & 0.0004 & 0.0491 & 0.0488 & 0.0130\\
H4$^\dag$ & 0.0036 & 0.018 & 0.088 & -1.2E-05 & 2.1E-04 & -- & 0.0004 & 0.0491 & 0.0488 & 0.0130 \\
H5&0.0032 & 0.016 & 0.09 & -1.3E-05 & 1.4E-04 & 9.1E-04 & -0.0001 & 0.0492 & 0.0493 & 0.0124\\
S1 &-- & -- & -- & 1.0E-04 & -- & -- & 0.0018 & 0.0366 & 0.0349 & 0.0133\\
S1$^{\dag}$&-- & -- & -- & 8.3E-05 & -- & -- & 0.0227 & 0.0359 & 0.0135 & 0.0151\\
S2 & -- & -- & -- & -- & 2.0E-04 & -- & 0.0004 & 0.0370 & 0.0366 & 0.0122\\
S3 & -- & -- & -- & -- & -- & 0.0015 & 0.0088 & 0.0370 & 0.0286 & 0.0121\\
S4& -- & -- & -- & -2.0E-05 & 2.3E-04 & -- & 0.0022 & 0.0370 & 0.0350 & 0.0121\\
S4$^{\dag}$& -- & -- & -- & -2.0E-05 & 2.1E-04 & -- & 0.0022 & 0.0370 & 0.0350 & 0.0121\\
S5 & -- & -- & -- & -1.3E-05 & 1.4E-04 & 9.0E-04 & 0.0003 & 0.0373 & 0.0369 & 0.0114\\
    \hline 
    \end{tabular}
    \vspace{0.2cm}
\begin{flushleft}
\footnotesize 
$^{(1)}$ A missing entry (denoted by ``--'') indicates that 
the corresponding tracer is not included in that case. 
See Sect.~\ref{sec_method} for definitions of the tracers listed in the table header, 
and for the units of the coefficients ($\alpha_s$) in the table body, as well as their conversion factors to the gas-to-dust ratio and $X_{\rm CO}$. 
The fit uncertainties, not including systematic calibration errors of the observations, are smaller than 1\%. \\
$^{(2)}$ A case marked with a superscript $^{\dag}$ indicates that 
a revised broad component, $I_{\rm HI,\ broad}^\dag = I_{\rm HI,\ broad}-I_{\rm HI,\ narrow}$, is used. \\
$^{(3)}$ The last three columns list the root mean square (rms) values of the fitted maps and of the residual map, in units of mg cm$^{-2}$. 
For reference, the rms of the dust map is 0.046 mg cm$^{-2}$.
\end{flushleft}
\end{table*}

\subsection{Planck thermal dust}
Thermal dust emission in the Polaris Flare is traced using all-sky maps from the \textit{Planck} mission \citep{2014A&A...571A..11P}. The satellite observed the sky in nine frequency bands from 30~GHz to 857~GHz, enabling the isolation of thermal dust emission from synchrotron, free--free, and cosmic microwave background components. For this study, we adopt the dust optical depth and temperature maps derived by the \textit{Planck} Collaboration \citep{2014A&A...571A..11P} from the high-frequency channels (353--857~GHz), which were modeled using the \citet[][DL07]{2007ApJ...657..810D} dust framework. The resulting maps have an effective angular resolution of 5\arcmin--7\arcmin. While coarser than the PPCOS CO data, this resolution is comparable to or finer than the EBHIS \ion{H}{I} dataset, enabling a direct spatial cross-comparison between the distinct ISM phases.
These \textit{Planck} data products are robust benchmarks for characterizing the structural, column density, and temperature properties of interstellar dust, as well as for estimating environmental gas-to-dust ratios \citep[e.g.,][]{2016A&A...596A.109P,2019A&A...622A..32L}. 


\section{Image overview}\label{over}
Prior to performing the linear decomposition, we examine the global spectra and morphological features of the Polaris Flare across the different gas and dust tracers.

\subsection{Global mean spectra}
The mean \ion{H}{I} spectrum of the Polaris Flare is well approximated by a two-component Gaussian model, comprising a narrow component (FWHM $\sim 7$~km\,s$^{-1}$) and a broad component (FWHM $\sim 27$~km\,s$^{-1}$), which are consistent with the cold and warm neutral media (CNM and WNM), respectively \citep[e.g.,][]{1982A&A...115..223M,2009ARA&A..47...27K}. The broad and narrow emissions span velocity ranges of $-40$ to $+20$~km\,s$^{-1}$ and $-8$ to $+5$~km\,s$^{-1}$, respectively. Centered at $-2.5$~km\,s$^{-1}$, the narrow \ion{H}{I} component aligns closely with the CO line profile (FWHM $\sim 3.9$~km\,s$^{-1}$), indicating that the molecular gas is embedded within this cold atomic phase \citep[e.g.,][]{2020A&A...639A..26K}. This narrow component is dominated by the CNM, which likely envelopes the CO-emitting regions and any associated CO-dark molecular gas \citep[e.g.,][]{2010ApJ...716.1191W,2016ApJ...819...22X,2022A&A...658A.140L}.
While the CNM, CO, and transitional gas phases collectively drive the observed dust emission \citep{2016ApJ...821..117K,2018A&A...619A..58K,2022ApJ...929...23H}, the diffuse broad-component WNM appears relatively dust-poor \citep[e.g.,][]{2023PhRvL.130p1002W}. These physical distinctions motivate a detailed spatial comparison between dust and multi-phase gas emissions within an isolated, structurally simple environment like the Polaris Flare. Consequently, we initiate our quantitative analysis by decomposing the \textit{Planck} dust map into a linear combination of integrated \ion{H}{I} and CO emissions.

\subsection{Images Overview}\label{sect_imoverview}
Figure~\ref{fig_overall} displays the spatial distributions of \textit{Planck} dust optical depth alongside various gas tracers, including integrated CO intensity maps ($I_{\rm ^{12}CO}$ and $I_{\rm ^{13}CO}$) and distinct \ion{H}{I} components ($I_{\rm HI,\ broad}$, $I_{\rm HI,\ narrow}$, and $T_{\rm HI,\ peak}$). The CO emission closely follows the densest, dust-bright ridges, whereas the \ion{H}{I} structures primarily trace the eastern boundary of the molecular cloud and more extended, diffuse environments (Figure~\ref{fig:marks}).
Morphological variations across the different \ion{H}{I} velocity components reflect distinct physical conditions. The broad component exhibits a highly smoothed distribution due to the line-of-sight blending of multiple velocity layers. Conversely, the narrow component displays fine-scale filamentary textures that match the striations seen in the high-resolution dust map, with both tracers clearly resolving the northern clump harboring the AN complex (Figure~\ref{fig:marks}).
Furthermore, the \ion{H}{I} channel map at $-2.5$~km~s$^{-1}$---corresponding to the global atomic spectrum peak---closely mimics the CO morphology. Notably, $T_{\rm HI,\ peak}$ (representing the peak channel intensity at $-2.5$~km~s$^{-1}$) highlights the AW complex (Figure~\ref{fig:marks}), which remains ambiguous in the integrated broad and narrow \ion{H}{I} maps. This structural difference confirms that the AW complex is denser, closer to the Galactic plane, and evolutionarily more mature than the AN complex, consisting predominantly of the cold neutral medium (CNM).
Based on these distinct spatial correlations, we reconstruct the thermal dust distribution using combinations of these multi-phase gas tracers via a multi-component linear inversion procedure detailed in Section~\ref{sec_method}.

\section{Linear fitting methodology}\label{sec_method}
In this section, we present the framework of multi-component linear decomposition (Section~\ref{sec_basic_fit}), interpret the physical meaning of the fitted parameters (Section~\ref{sec_phys_meaning}), and describe the regularization technique used to stabilize the fitting when evaluating highly multi-component datasets (Section~\ref{sec_reghook}).

\subsection{Basic linear fitting}\label{sec_basic_fit}
We denote the observed dust surface density map as $\mathcal{D}$ (the target map), expressed in units of mg~cm$^{-2}$ (where mg represents milligrams). Our objective is to reconstruct $\mathcal{D}$ using a linear combination of integrated intensity maps ($I_s$, in units of K~km~s$^{-1}$), where the subscript $s$ specifies the gas tracer. The molecular phase is traced by integrated intensity maps of $^{12}$CO, $^{13}$CO, and C$^{18}$O, while the atomic phase is traced by integrated 21~cm line maps obtained via discrete velocity integration or spectral decomposition. The modeled dust distribution is expressed as
\begin{equation}
    \mathcal{D}_{\rm fit} = \sum_{s=1}^{n+1} \alpha_s I_s,
\end{equation}
where $I_s$ encompasses all $n$ gas tracers, and $I_c \equiv I_{n+1} = 1$ everywhere represents a unity map that accounts for the large-scale background contribution. The corresponding coefficients $\alpha_s$ are free parameters determined through optimization, with the background coefficient denoted as $\alpha_c$. The fitted dust distribution excluding this background contribution is defined as
\begin{equation}
    \mathcal{D}_{\rm fit}^* = \sum_{s=1}^{n} \alpha_s I_s.
\end{equation}
The coefficients $\alpha_s$ are determined by minimizing the squared residuals between the observed and modeled dust maps over the spatial region covered by the CO observations (Figure~\ref{fig_overall}):
\begin{equation}
    \{ \alpha_s \} = \underset{\alpha_s}{\rm argmin} \; \left\| \mathcal{D} - \sum_{s=1}^{n+1} \alpha_s I_s \right\|^2. \label{eq_fitprocedure}
\end{equation}
The matrix-algebraic formulation of this least-squares solution is detailed in Appendix~\ref{app_matrix}. 
We solve this optimization problem using the exact matrix inversion routine \texttt{np.linalg.lstsq} from NumPy. 

For notational convenience when no ambiguity exists, we omit the explicit coefficients $\alpha_s$ and denote the weighted sum of tracers simply by their names or indices (e.g., $\mathcal{D}_{\rm fit} = {\rm ^{12}CO} + \ion{H}{I} + I_c$). Under this shorthand convention, the residual map $\mathcal{R}$ is defined as
\begin{equation}
    \mathcal{R} = \mathcal{D} - \mathcal{D}_{\rm fit} = \mathcal{D} - {\rm ^{12}CO} - \ion{H}{I} - I_c.
\end{equation}


\subsection{Physical meaning of fitted parameters}\label{sec_phys_meaning}
For atomic gas, $\alpha_{\rm HI}$ has units of mg~cm$^{-2}$~K$^{-1}$~km$^{-1}$~s. Because 21~cm emission in the low-density Polaris Flare remains optically thin, the integrated line intensity scales with the atomic hydrogen column density via \citep{2003ApJ...585..823L,2023ARA&A..61...19M}\footnote{For a Gaussian profile, $I \approx 1.06\,A\,\Delta V$, where $A$ is the peak and $\Delta V$ is the FWHM. If $I \approx A\,\Delta V$ is assumed, the conversion factor in Eq.~\ref{eq_HI_N} adjusts to $1.94\times 10^{18}$~cm$^{-2}$.}:
\begin{equation}
    N(\ion{H}{I}) = 1.84 \times 10^{18} \, \left( \frac{I_{\rm HI}}{\rm K\,km\,s^{-1}} \right) \ {\rm cm}^{-2}. \label{eq_HI_N}
\end{equation}
Adopting a mean atomic weight of 1.4 to account for a $\sim$40\% helium mass fraction relative to hydrogen, the corresponding atomic gas surface density is expressed as
\begin{equation}
    \Sigma_{\rm HI} = 4.3 \times 10^{-3} \, \left( \frac{I_{\rm HI}}{\rm K\,km\,s^{-1}} \right) \ {\rm mg}\,{\rm cm}^{-2}.
\end{equation}
Consequently, the gas-to-dust mass ratio ($a$) linked to $\alpha_{\rm HI}$ is given by
\begin{equation}
    a = \frac{4.3 \times 10^{-3}}{\alpha_{\rm HI}}. \label{eq_gasdustratio}
\end{equation}
A value of $\alpha_{\rm HI} = 4.3 \times 10^{-5}$ yields the canonical gas-to-dust mass ratio of $a \sim 100$, where larger coefficients imply a lower gas-to-dust ratio. For molecular gas, assuming a mean molecular weight of 2.8 and defining $a_{100} = a / 100$, the coefficient $\alpha_{\rm ^{12}CO}$ determines the CO-to-H$_2$ conversion factor $X_{\rm ^{12}CO}$ via
\begin{equation}
    X_{\rm ^{12}CO} = 2.14 \times 10^{22} \, a_{100} \, \alpha_{\rm ^{12}CO} \ {\rm cm}^{-2} \ ({\rm K\ km\ s^{-1}})^{-1}. \label{eq_Xco}
\end{equation}
A value of $\alpha_{\rm ^{12}CO} \sim 0.01$ reproduces the standard Galactic $X_{\rm ^{12}CO}$ factor of $2 \times 10^{20}$~cm$^{-2}$~(K~km~s$^{-1}$)$^{-1}$ \citep{2001ApJ...547..792D}.

\subsection{Regularized linear fitting}\label{sec_reghook}
When the predictor maps are highly correlated—a condition that frequently occurs when scaling to large numbers of components, such as decomposing individual velocity channels (Section~\ref{sec_fullspecfit})—the underlying correlation matrix becomes ill-conditioned. This ill-conditioning produces non-physical, high-frequency numerical oscillations in the solution vector $\alpha$. To suppress these instabilities, we employ Tikhonov (ridge) regularization \citep{tikhonov1977solutions}, which stabilizes the inversion by adding an identity matrix scaled by a regularization parameter $\beta$ to the correlation matrix (Section~\ref{sec_aboutregularization}).

\begin{figure}
    \centering
    \includegraphics[width=0.99\linewidth]{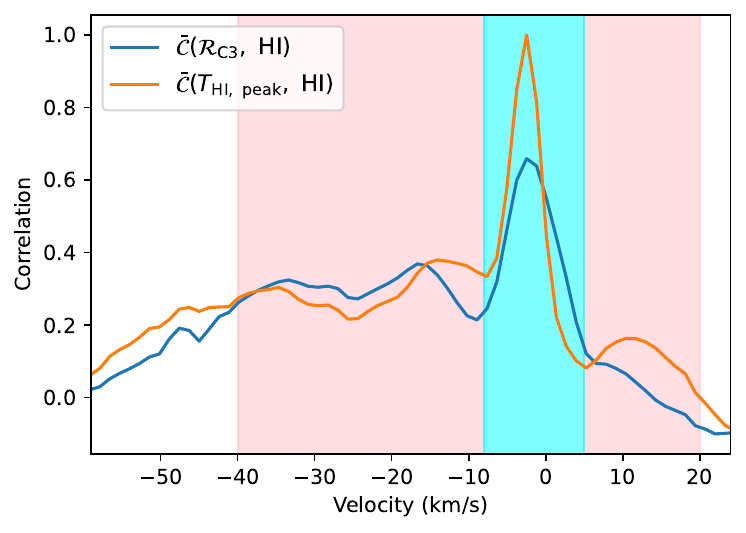}
    \caption{Detrended correlation (Sect. \ref{sec_detrendcross}) of the \ion{H}{I} channel maps with $\mathcal{R}_{\rm C3}$ (blue line) and the $-2.5$ km\,s$^{-1}$ channel map (orange line). }
    \label{fig:correlationcurve}
\end{figure}

\begin{figure*}
    \centering
    \includegraphics[width=0.98\linewidth]{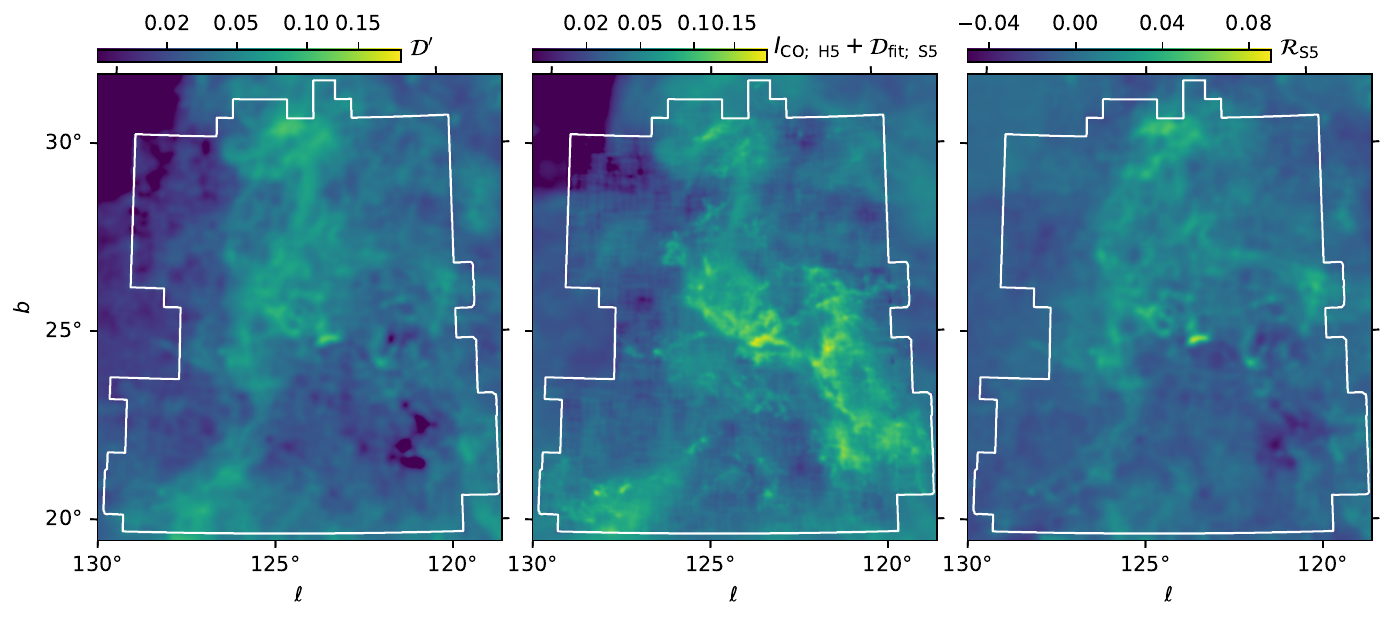}
\caption{
Left: CO residual image from the H5 fitting, used as the objective map for the S1--S5 fittings (Sect.~\ref{sec_smfit}). 
Middle: Best-fit map, obtained by combining the CO contribution from H5 with the \ion{H}{I} contribution from S5. 
Right: Residual map of case~S5. 
Note that the fitted maps and residual map share the same color scales as the corresponding panels in Fig.~\ref{fig_fitting}.
}
\label{fig:bestfig}
\end{figure*}

\begin{figure}
    \centering
    \includegraphics[width=0.95\linewidth]{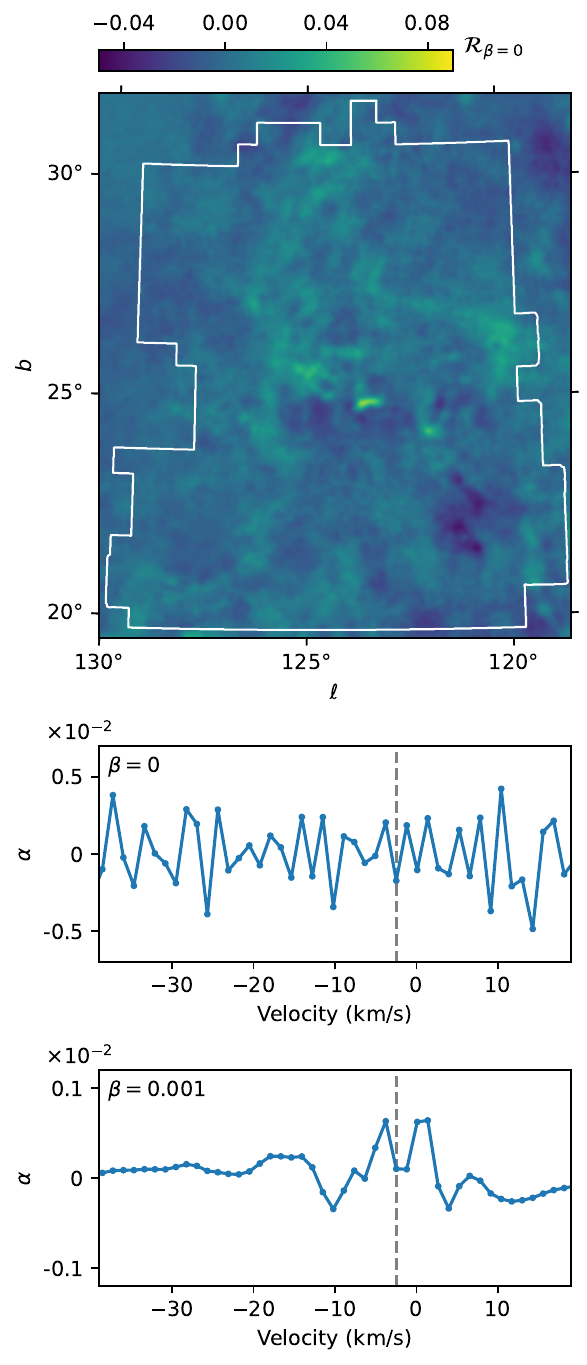}
    \caption{
Upper: Residual map from the full-spectrum fitting, using all \ion{H}{I} channels within $-40$ to $+20$~km~s$^{-1}$. 
Middle: Fitted $\alpha$ values corresponding to the full-spectrum fitting ($\beta=0$). 
Lower: Same as the middle panel but with a regularization parameter $\beta=0.001$
(Sect. \ref{sec_fullspecfit}).
}
    \label{fig:radialfitting}
\end{figure}

\begin{figure*}
    \centering
    \includegraphics[width=0.99\linewidth]{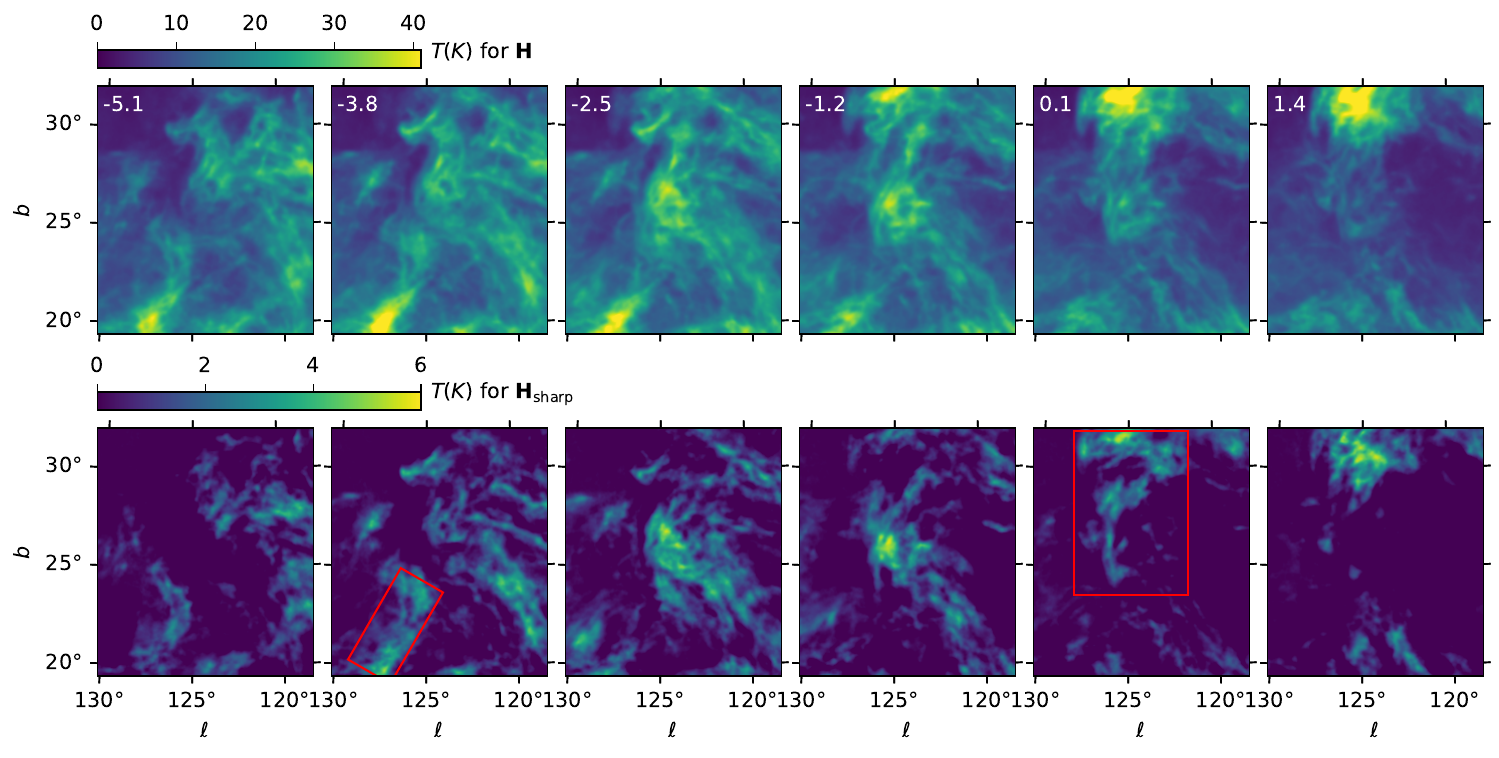}
\caption{
Channel maps of the original \ion{H}{I} cube ($\mathcal{H}$; upper panels)
and the spectrally sharpened, positive-thresholded cube ($\mathcal{H}_{\rm sharp}'$; lower panels; Sect.~\ref{sec_reg_phy}).
The velocity is indicated in the upper-left corner of each panel.
The upper and lower panels each share a separate colorbar, displayed at the top of the first-column panels.
The red boxes mark the southern and northern parts of the eastern wedge (Sect.~\ref{sect_imoverview}), 
which stand out clearly in the channel maps of $\mathcal{H}_{\rm sharp}$.
\label{fig_hsharp}
}
\end{figure*}

\begin{figure}
    \centering
    \includegraphics[width=0.9\linewidth]{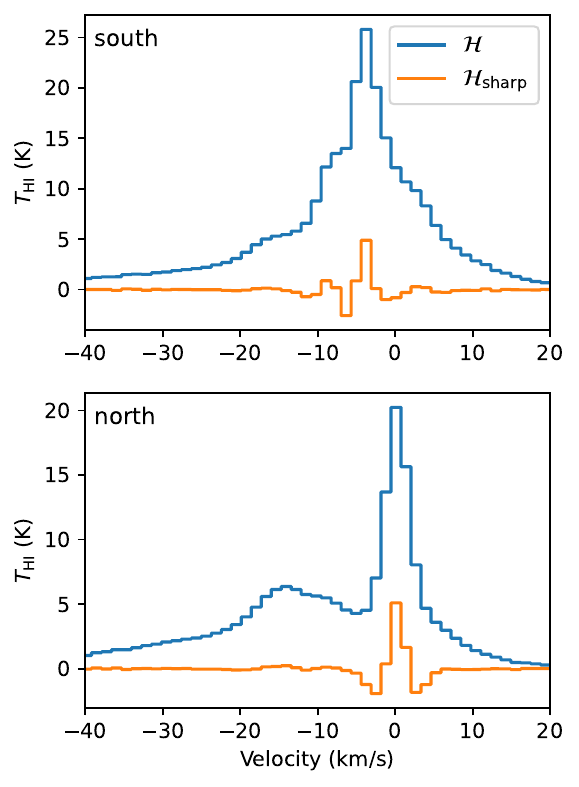}
    \caption{Typical \ion{H}{I} spectra in the southern (upper) and northern (lower) parts of the eastern edge, as marked by the red boxes in Figure~\ref{fig_hsharp}. 
The blue lines show the original spectra, while the orange lines show the spectrally sharpened spectra (Sect.~\ref{sec_reg_phy}).}
    \label{fig:typicalnarrowlines}
\end{figure}

\section{Decomposition by moment 0 maps} \label{sec_basicdecomp}
In this section, we apply the basic linear fitting (Sect. \ref{sec_basic_fit}) by using the moment-0 maps of CO and \ion{H}{I} as dust tracers. The fitting results are summarized in Table~\ref{tab_fit}.
\subsection{CO-only cases}
For the fitting, the CO maps are smoothed to match the spatial resolution of the dust map (with a spatial resolution of $\sim5\arcmin$).

\subsubsection{Fitting results of CO-only cases}
When only $I_{\rm ^{12}CO}$ is adopted (Case~C1 in Table~\ref{tab_fit}), the fitting yields a coefficient $\alpha_{\rm ^{12}CO} = 0.0055$. The fitted map, with the constant component excluded ($\mathcal{D}_{\rm C1}^*=\mathcal{D}_{\rm C1}-I_c$), has a root mean square of 0.0212~mg~cm$^{-2}$. The derived $\alpha_{\rm ^{12}CO}$ corresponds to an $X_{\rm ^{12}CO}$ value of $1.2\times 10^{20}$~cm$^{-2}$~K$^{-1}$~km$^{-1}$~s (Eq.~\ref{eq_Xco}), slightly lower than the commonly adopted value of $2\times 10^{20}$~cm$^{-2}$~K$^{-1}$~km$^{-1}$~s, but broadly consistent with the lower limit of the fiducial range ($(1\text{--}5)\times 10^{20}$~cm$^{-2}$~K$^{-1}$~km$^{-1}$~s) in observations \citep[e.g.,][]{2013ARA&A..51..207B}.
Only about 40\% ($0.0212/0.05$) of the dust surface density can be reproduced by $^{12}$CO alone (here 0.05 is the RMS of $\mathcal{D}$). A straightforward explanation might be that this low value of $X_{\rm ^{12}CO}$ is an artifact of the unmodeled dust; however, this is excluded by the fact that this residual dust is actually associated with the atomic component and is mostly recovered by \ion{H}{I} emission (Section~\ref{sec_HIfit}). Therefore, a more thorough physical explanation is required. For gas at a constant temperature, sub-thermal under-excitation at low densities and self-absorption at high densities both lead to an increase in $\alpha_{\rm ^{12}CO}$, meaning a natural lower bound for $\alpha_{\rm ^{12}CO}$ exists at intermediate densities conceptually between the critical densities of excitation and self-absorption. Our low value implies that the $^{12}$CO-traced gas in the Polaris Flare corresponds precisely to this intermediate density range. This is highly consistent with the fact that the Polaris Flare is a diffuse cloud environment that fundamentally lacks the dense, heavily self-absorbing structures typical of giant molecular clouds in the Galactic plane \citep{2015ARA&A..53..583H}.

Including $^{13}$CO in the decomposition (Case~C2) reduces $\alpha_{\rm ^{12}CO}$ to 0.0041 while yielding $\alpha_{\rm ^{13}CO} = 0.017$. The resulting modeled map ($\mathcal{D}_{\rm C2}$) exhibits finer small-scale structures compared to $\mathcal{D}_{\rm C1}$, though its global power (quantified by the rms of $\mathcal{D}_{\rm C2}$) increases marginally (Table~\ref{tab_fit}). The derived coefficient ratio satisfies
\begin{equation}
    \alpha_{\rm ^{13}CO}/\alpha_{\rm ^{12}CO} \sim 4, \label{1213alpharatio}
\end{equation} 
sitting far below the canonical $^{12}$C/$^{13}$C isotopic ratio of $\sim 70$ in the local ISM. This offset is expected since the extensive, diffuse peripheral molecular gas is traced predominantly by $^{12}$CO emission.
Within the localized $^{13}$CO emission regions, the integrated intensity ratio between $^{12}$CO and $^{13}$CO ($I_{12;13}$) ranges from 5 to 25 (Paper~I). Combined with Eq.~\ref{1213alpharatio}, the values of $I_{12;13}$ imply a ratio of gas mass traced by $^{13}$CO relative to $^{12}$CO of $\sim$0.2 to $\sim$1. Consequently, in regions characterized by the brightest $^{13}$CO emission, the surface density of gas traced by $^{13}$CO becomes directly comparable to that of the surrounding peripheral gas traced by $^{12}$CO.

The emission region of C$^{18}$O is localized to a few highly compact areas, such as the densest part of the Polaris cloud\footnote{The Polaris cloud is a small, bright region within the Polaris Flare, covering $\sim 0.6$~deg$^2$ \citep[][see also Paper~I of the PMO Polaris CO survey]{2001A&A...366..636B}.}. When $I_{\rm C^{18}O}$ is included in the decomposition (Case~C3), the derived $\alpha_{\rm ^{12}CO}$ coefficient and the global rms residual of $\mathcal{D}_{\rm fit}$ remain nearly unchanged. The resulting coefficient ratio yields $\alpha_{\rm C^{18}O}/\alpha_{\rm ^{13}CO} = 5.3$. Within the brightest C$^{18}$O emission structures, the integrated line intensity ratio between $^{13}$CO and C$^{18}$O is $\sim$6.
Combined with the coefficient ratio, this intensity ratio indicates that the gas surface density traced by C$^{18}$O is directly comparable to that of the surrounding gas layer traced by $^{13}$CO. Similar to the $^{12}$CO--$^{13}$CO boundary, this self-similar mass allocation suggests that the spatial architecture of the molecular components within the Polaris Flare closely follows a power-law density--radius relation of $\rho \propto R^{-1}$, consistent with a constant column density profile spanning the hierarchical cloud structures \citep{1981MNRAS.194..809L,2010A&A...519L...7L}.


Across all three molecular fitting configurations (Cases~C1--C3), the derived CO-associated dust component ($\mathcal{D}_{\rm C}^*$) remains nearly invariant, accounting for a stable fraction of 44\% (Table~\ref{tab:dustfractions}). This indicates that less than half of the total thermal dust emission in the Polaris Flare is traced by CO emission alone. Consequently, these results underscore that the clear majority of the dust mass resides outside the CO-emitting molecular phase, and must instead be linked to atomic gas or transitional dark components. This shortfall provides the primary physical motivation for incorporating \ion{H}{I} into the subsequent decomposition frameworks.

\subsubsection{Cross-correlation of dust residuals and \ion{H}{I}}\label{sec_detrendcross}
In the CO-only configuration (Case~C1--C3), the resulting residual map ($\mathcal{R}_{\rm C3}$) reveals two structural regimes: a sharp eastern boundary bordering the dust-bright region and an extended, diffuse envelope (Figure~\ref{fig_fitting}). This eastern boundary divides into three distinct features: a northern clump containing the AN complex, an S-shaped twisted filament running north--south, and a C-shaped structure enclosing the AC and AW complexes (the main C-shape). A secondary, lower-intensity C-shaped arc is also visible to the north (the northern C-shape). These morphological features can be broadly identified across different \ion{H}{I} velocity components (Section~\ref{sect_imoverview}). Consequently, we aim to isolate the specific \ion{H}{I} emission that dictate these distinct dust residual structures.

Prior to incorporating atomic emission into the global linear optimization, we examine the channel-by-channel correlation between $\mathcal{R}_{\rm C3}$ and the \ion{H}{I} data cube. To maximize sensitivity to morphological alignment while removing baseline intensity scales, we calculate the detrended correlation coefficient. For any two spatial maps $M_1$ and $M_2$, this metric is defined as
\begin{equation}
    \bar{\mathcal{C}}(M_1, M_2) = \frac{\sum_j \bar{M}_{1,j}\,\bar{M}_{2,j}}{\sqrt{\left(\sum_j \bar{M}_{1,j}^2\right)\left(\sum_j \bar{M}_{2,j}^2\right)}},
\end{equation}
where $j$ represents the pixel index and $\bar{M} = M - \langle M \rangle$ denotes the mean-subtracted spatial template. This coefficient quantifies the structural coherence between the two datasets, where values approaching unity indicate strict morphological similarity. Applying this estimator across the \ion{H}{I} velocity channels isolates the specific atomic velocity intervals that are dynamically linked to the dust structures unprobed by molecular CO.

Figure~\ref{fig:correlationcurve} displays this channel-resolved detrended correlation between $\mathcal{R}_{\rm C3}$ and the \ion{H}{I} cube. The unmodeled dust residual correlates most strongly with atomic channels spanning $-8$ to $+5$~km~s$^{-1}$ (cyan shading in Figure~\ref{fig:correlationcurve}), matching the narrow atomic velocity component that matches the systemic molecular velocity (Figure~\ref{fig:meanspec_COHI}). The peak correlation ($\bar{\mathcal{C}} \sim 0.6$) occurs precisely at $-2.5$~km~s$^{-1}$, aligning with the global atomic spectral peak. Conversely, channels tracking the broad velocity component between $-40$ and $-8$~km~s$^{-1}$ (pink shading in Figure~\ref{fig:correlationcurve}) exhibit a suppressed, flat correlation plateau of $\sim 0.3$, while channels exceeding $+5$~km~s$^{-1}$ show no statistically meaningful structural correlation.

Physically, we interpret the broad component as diffuse WNM, the narrow component as bulk CNM blended with a warm envelope, and the $-2.5$~km~s$^{-1}$ peak channel as localized CNM tracing the atomic-to-molecular interface. The correlation ratio between this peak channel and the WNM channels indicates that up to one-third of the interface CNM column density is kinematically blended with background WNM emission. Across the entire bulk CNM interval ($-8$ to $+5$~km~s$^{-1}$), this blending fraction increases to approximately one-half, consistent with the mean profiles shown in Figure~\ref{fig:meanspec_COHI}. Consequently, the weak correlation observed between the WNM channels and the residual dust map is likely an indirect artifact driven by spatial coupling between the warm and cold neutral phases, implying that the WNM itself is largely dust-poor.

\begin{table}[htbp]
\centering
\caption{Dust mass fractions for C3, H5, and final fittings.}
\label{tab:dustfractions}
\begin{tabular}{lccc}
\hline\hline
Component & C3$^{(1)}$ & H5 & Final \\
\hline
CO-dust fraction          & 22\,\%$^{(2)}$ & 20\,\% & 20\,\% \\
\ion{H}{I}-dust fraction  & --              & 70\,\% & 76\,\% \\
Residual dust fraction    & 10\,\%          & 10\,\% & 4\,\% \\
Eastern ridge residual fraction$^{(3)}$ & 30\,\% & 25\,\% & 10\,\% \\
\hline
\end{tabular}\\
\vspace{2mm}
\raggedright \textit{Notes.} 
$^{(1)}$~In the C3 fitting, most dust is absorbed into the constant component.
$^{(2)}$~The CO-dust fraction would rise to 40\,\% if the constant component were excluded during fitting.
$^{(3)}$~Ratio of mean dust surface density of the eastern edge, between residual and raw dust maps.
\end{table}

\begin{figure*}[t]
    \centering
    \includegraphics[width=0.9\linewidth]{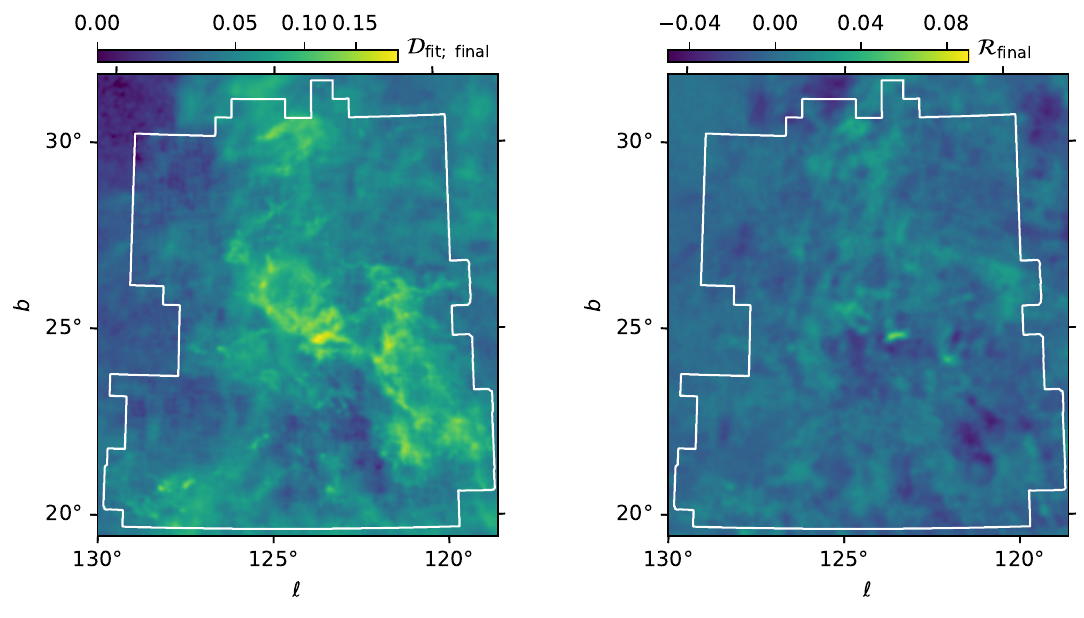}
    \caption{Fitted dust map, $\mathcal{D}_{\rm fit;\ final}$ (Eq.~\ref{eq_Dfinal}), and residual map, $\mathcal{R}_{\rm final}$, from the final linear fitting that includes CO moment-0 maps, \ion{H}{I} channel maps, and the spectrally sharpened \ion{H}{I} emission (Sect.~\ref{sec_reg_phy}). 
\label{fig_finalfit}}
\end{figure*}

\subsection{Fitting cases including \ion{H}{I}}\label{sec_HIfit}

For the fittings in this section, the CO maps are smoothed to match the spatial resolution of the dust map ($\sim5\arcmin$), while the \ion{H}{I} maps are left at their native resolution (10.8\arcmin) due to the coarser beam.

\subsubsection{Broad \ion{H}{I} case}
In Case~H1, we incorporate the integrated broad \ion{H}{I} emission alongside the molecular CO maps. The root-mean-square (rms) residual of the resulting map decreases from 0.016 in the CO-only configuration (Case~C3) to 0.013 (Table~\ref{tab_fit}), while the constant background component $I_c$ drops from 0.034 to nearly zero. This suppression demonstrates that diffuse, large-scale dust distributions are structurally well accounted for by the extended atomic gas emission. The optimization yields a fitting coefficient of $\alpha_{\rm HI, broad} = 1.0 \times 10^{-4}$ which, via Eq.~\ref{eq_gasdustratio}, translates to a gas-to-dust mass ratio of $a \sim 43$ compared to the canonical value of 100. This low gas-to-dust ratio points to a significant dust component that correlates spatially with the atomic gas template but remains untraced by CO or integrated broad \ion{H}{I} alone.
Conversely, utilizing a modified broad atomic component defined by subtracting the narrow spectral line profile (Case~H1$^\dag$) yields a poorer overall optimization characterized by an elevated rms residual and an expanded $I_c$ background contribution (Table~\ref{tab_fit}). This performance divergence confirms that the diffuse warm neutral medium (WNM) is a minor direct contributor to the total dust budget.


\subsubsection{Narrow \ion{H}{I} included cases}
In Case~H2, we incorporate the integrated narrow \ion{H}{I} emission ($-8$ to $+5$~km~s$^{-1}$) alongside the CO maps. The root-mean-square (rms) residual decreases to 0.013 (Table~\ref{tab_fit}), slightly lower than in the broad-\ion{H}{I} configuration (Case~H1), while $I_c$ remains near zero. The optimization yields $\alpha_{\rm HI, narrow} = 2.0 \times 10^{-4}$ which, via Eq.~\ref{eq_gasdustratio}, translates to a low gas-to-dust mass ratio of $a \sim 22$, indicating a high dust content associated with this component. This implies that dust traced by narrow atomic emission is predominantly associated with the dense cold neutral medium (CNM) at the atomic--molecular interface. This depressed gas-to-dust ratio is naturally explained by the presence of CO-dark molecular gas, which contributes to the thermal dust emission but lacks corresponding CO line emission.

In Case~H4, both the narrow and broad \ion{H}{I} components are fitted simultaneously with the molecular maps. The resulting rms residual and $I_c$ background remain comparable to Case~H2 (Table~\ref{tab_fit}), confirming that the narrow component dominates the atomic dust contribution. For Case~H4$^\dag$, where the broad component template is adjusted by subtracting the narrow emission, the reconstructed map ($\mathcal{D}_{\rm fit}$) and rms residual are identical to Case~H4. This behavior is expected because the basis pairs ($I_{\rm HI, broad}$, $I_{\rm HI, narrow}$) and ($I_{\rm HI, broad}-I_{\rm HI, narrow}$, $I_{\rm HI, narrow}$) span the same linear subspace (Appendix~\ref{app_matrix}). The stable behavior of these coefficients further confirms that the diffuse WNM layer contributes negligibly to the global dust mass.

\subsubsection{Peak \ion{H}{I} included cases}
In Case~H3, we incorporate only the peak \ion{H}{I} channel (centered at $-2.5$~km~s$^{-1}$) alongside the CO maps. The root-mean-square (rms) residual decreases marginally to 0.0130, while the constant component $I_c$ remains at 0.0080 (Table~\ref{tab_fit}), demonstrating that the peak channel alone is insufficient to reconstruct the bulk dust distribution. However, this configuration successfully reproduces the northern C-shaped feature that remains uncaptured by the standard integrated narrow or broad \ion{H}{I} templates. Introducing this peak channel template also induces a modest decrease in the derived $\alpha_{\rm CO}$ parameters, which is physically consistent with a scenario where this ultra-narrow cold atomic phase—the very cold neutral medium (VCNM)—is spatially co-located with the CO-emitting molecular structures.

In Case~H5, all three atomic templates (integrated narrow, integrated broad, and peak channel) are optimized simultaneously with the molecular maps. This multi-component combination minimizes the rms residual to 0.0124 and drives $I_c$ to $-0.0001$ (Table~\ref{tab_fit}), significantly improving the reconstruction of the global thermal dust map. Under this joint framework, the narrow component coefficient stabilizes at $\alpha_{\rm HI, narrow} = 1.4 \times 10^{-4}$ while $\alpha_{\rm CO}$ drops relative to configurations lacking these velocity-resolved atomic constraints. These variations confirm that the dust excess originates predominantly within the densest atomic gas tracing the atomic--molecular interface (the interface CNM or VCNM) and within dark molecular hydrogen gas unprobed by 21~cm line emission. Ultimately, this joint inversion yields a self-consistent dust map decomposition that isolates the dominant contributions of cold atomic and molecular gas phases while confirming that the diffuse WNM is dust-poor.

\subsubsection{Dust fraction of H5 fitting}
Along the eastern edge, the mean value of the dust residual is 
0.02\,mg\,cm$^{-2}$, corresponding to about 25\% of the value in the 
original dust map (0.08\,mg\,cm$^{-2}$). The total residual dust mass, 
assuming a distance of 150\,pc, is $\sim$13\,$M_\odot$, compared to 
$\sim$130\,$M_\odot$ for the total dust. This implies that the residual 
component contributes about 10\% of the overall dust mass in the Polaris 
region. The majority of the dust is instead associated with CO (20\%) 
and with \ion{H}{I} (70\%). 
These fractions are summarized in Table~\ref{tab:dustfractions}, which 
provides a compact comparison between the H5 fitting and the final 
fitting (Sect. \ref{sec_reg_phy}).

\begin{figure*}[t]
    \centering
    \includegraphics[width=0.99\linewidth]{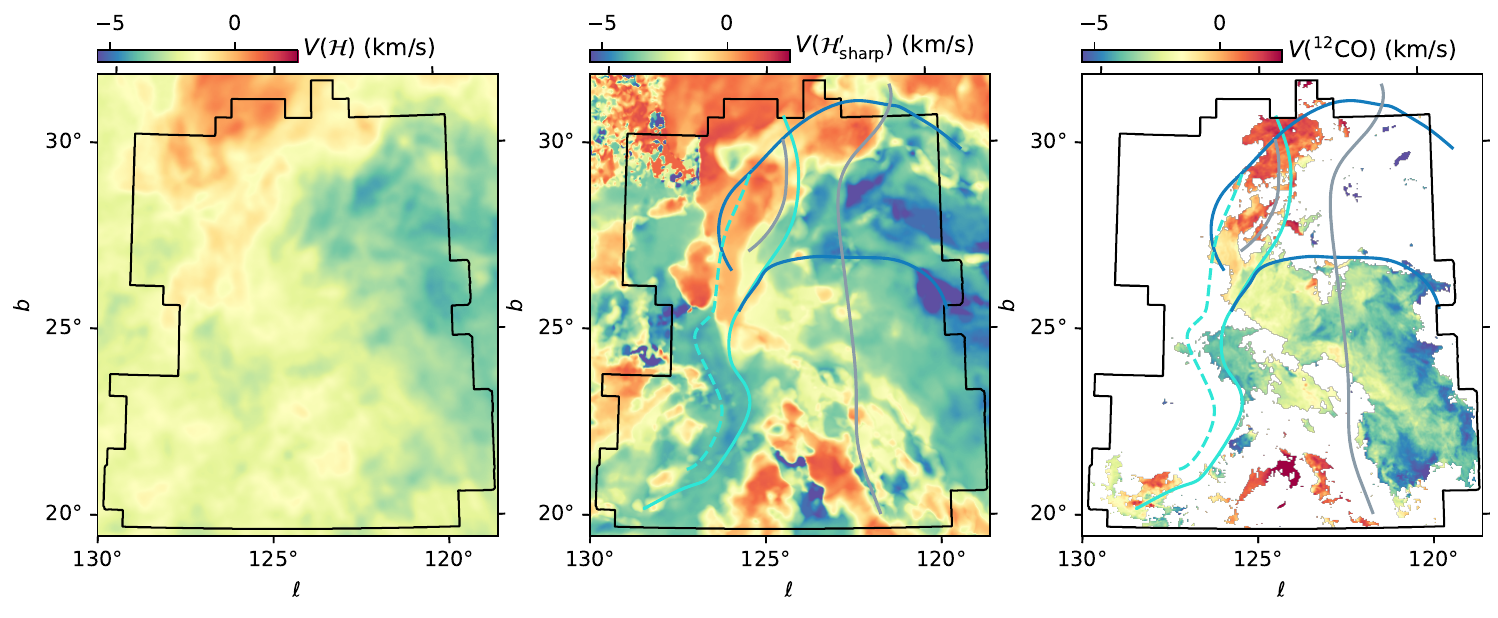}
    \caption{
Left: Moment-1 map of \ion{H}{I} ($\mathcal{H}$) over the velocity range of the narrow component (-8 to 5 km s$^{-1}$). 
Middle: Moment-1 map of the spectrally sharpened and positively thresholded \ion{H}{I} cube ($\mathcal{H}_{\rm sharp}'$; Sect. \ref{sec_reg_phy}), weighted by $\alpha_{\beta=0.001}$ (see Sect. \ref{sec_findnarrowspec} and Figure \ref{fig:radialfitting}), over the same velocity range. 
Right: Moment-1 map of $^{12}$CO from Paper I. 
All three panels share the same color scale. 
The curves shown in the middle and right panels have the same meaning as in Figure \ref{fig:marks}; their colors differ for clarity but can be recognized by their shapes.
\label{fig_velos}
}
\end{figure*}

\subsection{Fitting configurations with matched resolution}\label{sec_smfit}
Spatial resolution mismatches between the thermal dust maps and the coarse \ion{H}{I} data can introduce significant uncertainties (up to a factor of $\sim2$) in the derived fitting coefficients under extreme structural gradients (Section~\ref{sec_theoryaboutbeameffect}). To suppress these beam-mismatch artifacts, we smooth the molecular-subtracted dust template from the Case~H5 inversion to the native 10.8\arcmin\ spatial resolution of the EBHIS \ion{H}{I} dataset. We define this resolution-matched target map $\mathcal{D}'$ as
\begin{equation}
    \mathcal{D}' = \langle \mathcal{D} - I_{\rm CO;\ H5} \rangle_{\rm 10.8'}, \label{eq_Dsm}
\end{equation}
where $\langle \cdot \rangle_{\rm 10.8'}$ denotes a 10.8\arcmin\ Gaussian smoothing convolution, and $I_{\rm CO;\ H5} = \sum \alpha_{s,\rm H5} I_s$ accounts for the combined contributions of all three CO isotopologues. We then utilize $\mathcal{D}'$ as the target map (Section~\ref{sec_basic_fit}) to perform a linear fit using only the \ion{H}{I} emission. This framework ensures that the derived $\alpha$ parameters are unaffected by beam-smearing biases.

The resolution-matched optimization reveals that the broad \ion{H}{I} coefficient remains small, the narrow \ion{H}{I} coefficient is essentially unchanged, and the peak channel coefficient decreases by no more than $\sim 10\%$. This stability indicates that the beam effect is negligible for our analysis. Methodologically, the impact of the beam depends on the ratio between the effective resolution---defined by the beam size plus the characteristic width of the emission features---and the physical spatial scale of the structures under consideration. Hence, these stable results are expected and primarily reflect that the peak channel traces intrinsically finer structures than the broader CNM component.
Meanwhile, the remaining $\sim 25\%$ of the dust along the eastern edge (Table~\ref{tab:dustfractions}) that cannot be reproduced by either CO or \ion{H}{I} emission is likely associated with CO-dark molecular gas. This represents transitional regions where H$_2$ has already formed but CO is either not yet abundant or not efficiently excited.

\section{Full-spectrum fitting}\label{sec_fullspecfit}
In Section~\ref{sec_basicdecomp}, we demonstrated that \ion{H}{I}-associated dust concentrates predominantly within the CNM ($\Delta V < 10$~km\,s$^{-1}$), particularly within the coldest atomic structures tracing the atomic-to-molecular interface. This phase exhibits line widths comparable to, or even narrower than, the native velocity resolution of the EBHIS survey. However, the exact kinematics of these coldest gas components remain poorly constrained. While Section~\ref{sec_basicdecomp} establishes that the molecular-subtracted dust residual correlates most strongly with the peak channel at $-2.5$~km\,s$^{-1}$ on global scales, it remains unclear whether this spatial correlation persists uniformly across distinct regions of the Polaris Flare. Furthermore, it is uncertain whether the unmodeled dust emission can be better reproduced by incorporating a combination of multiple adjacent channels. Another unresolved issue is whether the significant residual dust component—which persists even after accounting for the integrated broad and narrow \ion{H}{I} contributions—can be physically accounted for by atomic emission originating from specific, individual velocity channels.

\subsection{Standard fitting}\label{sec_dospecfit}
To isolate the atomic gas channels contributing to the dust, we perform full-spectrum fitting where each channel map of the \ion{H}{I} cube (from $-60$ to $+30$~km\,s$^{-1}$) serves as an individual predictor $I_s$, yielding $n \sim 60$. At the 10.8\arcmin\ EBHIS resolution, the dust target map retains $N \sim 3600$ independent spatial resolution elements, ensuring the regression is well-constrained since $N \gg n$. However, because the physical line widths of the diffuse WNM and bulk CNM span multiple channels, the predictor matrix becomes highly ill-conditioned (Section~\ref{sec_aboutregularization}).

Applying the fitting to $\mathcal{D}'$ (Section~\ref{sec_smfit}) without regularization ($\beta=0$), the spatial residual map (upper panel of Figure~\ref{fig:radialfitting}) mimics the Case~H5 residual, indicating that the remaining unmodeled dust emission cannot be adequately traced by standard CO or \ion{H}{I} emission despite the expanded parameters. Crucially, the derived coefficients $\alpha_s$ exhibit severe numerical oscillations across channels (middle panel of Figure~\ref{fig:radialfitting}), preventing any meaningful physical interpretation of the velocity-dependent results.

\subsubsection{Very narrow components by regularized fitting}\label{sec_findnarrowspec}
To suppress these numerical oscillations, we apply Tikhonov regularization (Sections~\ref{sec_reghook} and \ref{sec_aboutregularization}) using a fiducial value of $\beta = 0.001$, which leaves the spatial residual map virtually unchanged. Crucially, the velocity dependence of $\alpha_s$ becomes well-behaved, revealing two distinct kinematic peaks centered at $-3.8$~km\,s$^{-1}$ and $0$~km\,s$^{-1}$ (lower panel of Figure~\ref{fig:radialfitting}). Remarkably, the regularized solution does not peak at $-2.5$~km\,s$^{-1}$, which is the individual channel exhibiting the highest isolated correlation with the target map $\mathcal{D}'$. This behavior may either (1) reflect residual ill-conditioning within the correlation matrix, or (2) physically trace actual \ion{H}{I} structures at blueshifted ($-3.8$~km\,s$^{-1}$) and redshifted ($0$~km\,s$^{-1}$) velocities. Given the large-scale velocity gradient across this region (Paper~I), we favor the latter physical scenario. This implies that the dust-associated atomic gas structures must possess very narrow line widths comparable to the channel resolution itself.

\begin{figure}[t]
    \centering
    \includegraphics[width=0.99\linewidth]{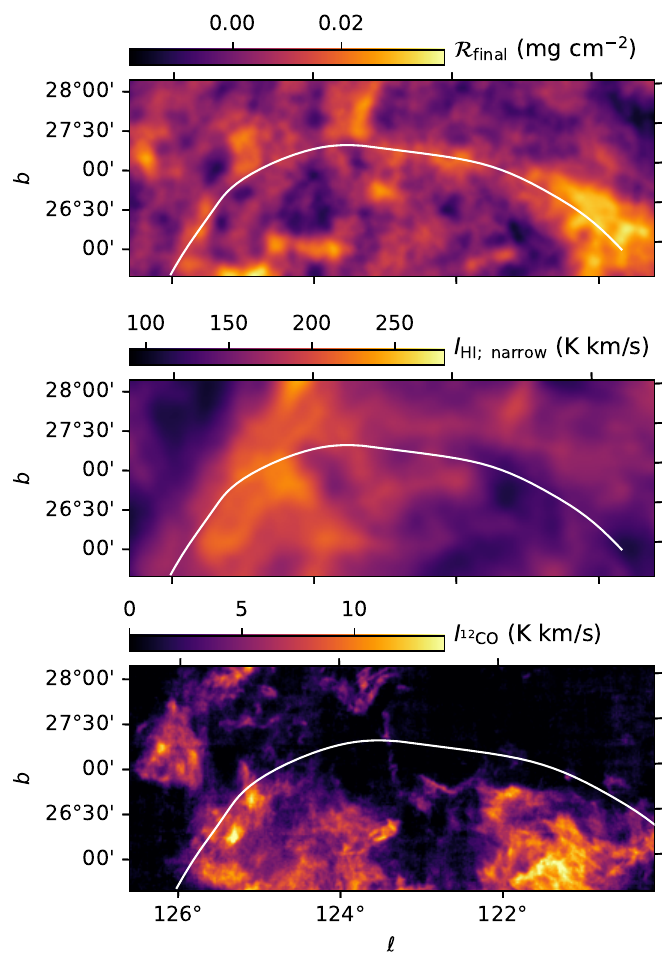}
    \caption{Images of the main C-shaped feature (Figure~\ref{fig:marks}). 
Top: Dust residuals from the final fitting (Sect.~\ref{sec_thefinalfit}). 
Middle: Moment-0 map of the narrow \ion{H}{I} component (Sect.~\ref{over}). 
Bottom: Moment-0 map of $^{12}$CO (see also Figure~\ref{fig_overall}). 
The white line indicates the main C-shape structure (corresponding to the lower green line in Figure~\ref{fig:marks}).}
    \label{fig:mainC}
\end{figure}

\section{Locating the very narrow \ion{H}{I} components} \label{sec_reg_phy}
To investigate the spatial distribution of the very narrow velocity components that contribute to the dust emission (see Sect.~\ref{sec_findnarrowspec}), we apply the following decoupling method. 
We first smooth the \ion{H}{I} data cube ($\mathcal{H}$) along the spectral axis using a Gaussian kernel with a full width at half maximum (FWHM) of 3 pixels, producing a smoothed cube $\mathcal{H}_{\rm smooth}$. The difference between the original and smoothed cubes is then computed as
\begin{equation}
    \mathcal{H}_{\rm sharp} = \mathcal{H} - \mathcal{H}_{\rm smooth}.
\end{equation}
This method can be considered a high-pass filter and has previously been implemented for multiscale decomposition of astronomical maps \citep{2022ApJS..259...59L}. By applying this concept along the spectral axis, it effectively isolates the very narrow velocity components, allowing us to examine their spatial distribution across the Polaris Flare.

Figure~\ref{fig_hsharp} shows a comparison between the channel maps
of $\mathcal{H}$ and $\mathcal{H}_{\rm sharp}$ in the velocity range from $-5$ to $2$ km\,s$^{-1}$.
The $\mathcal{H}$ channel maps appear overall similar across different velocities,
whereas the $\mathcal{H}_{\rm sharp}$ maps vary much more dramatically.
Interestingly, the $\mathcal{H}_{\rm sharp}$ channel maps at $-3.8$ km\,s$^{-1}$
and $0.1$ km\,s$^{-1}$ clearly trace the southern and northern parts of the eastern edge,
which cannot be fitted simultaneously in the previous analysis.
These two channels correspond exactly to the two peak channels of $\alpha_{\beta=0.001}$
(see Sect.~\ref{sec_fullspecfit} and Fig.~\ref{fig:radialfitting}).

\subsection{Final linear fitting}\label{sec_thefinalfit}
To include the channel maps of $\mathcal{H}_{\rm sharp}$ in the dust map fitting, we slightly modify $\mathcal{H}_{\rm sharp}$. 
Specifically, we define a negative-value–clipped cube as
\begin{equation}
    \mathcal{H}_{\rm sharp}'(j) =
    \begin{cases}
        \mathcal{H}_{\rm sharp}(j), & \text{if } \mathcal{H}_{\rm sharp}(j) > 0, \\
        0, & \text{otherwise}.
    \end{cases}
\end{equation}
Including $\mathcal{H}_{\rm sharp}$ maps in the fitting does not provide additional information compared to the full-spectrum approach (Sect.~\ref{sec_fullspecfit}), since they are linear combinations of the $\mathcal{H}$ maps.  

Nevertheless, by adding the six $\mathcal{H}_{\rm sharp}'$ images shown in the lower panels of Fig.~\ref{fig_hsharp} to the full-spectrum fitting of Sect.~\ref{sec_dospecfit} ($\beta=0$), we obtain the final linear fit to the dust map, $\mathcal{D}_{\rm final}$, along with its residual (Fig.~\ref{fig_finalfit}).  
Here, we have also added back the CO component from the H5 fitting,
\begin{equation}
    \mathcal{D}_{\rm fit;\ final} = I_{\rm final;\ HI} + I_{\rm H5;\ CO}.
    \label{eq_Dfinal}
\end{equation}
The rms of the residual decreases to 0.008, lower than that of the pure full-spectrum fitting. This suggests that the dust not traced by either \ion{H}{I} or CO is less abundant than indicated by the moment-0 map fittings, e.g., the S5 fitting (Sect.~\ref{sec_smfit} and Table~\ref{fig_fitting}).  
Since this improvement arises from the increased number of free parameters, we treat $\mathcal{R}_{\rm final}$ as a lower limit of the CO-dark molecular gas, and $\mathcal{R}_{\rm S5}$ as an upper limit.

The image of $\mathcal{D}_{\rm fit;\ final}$, in comparison with the fitted images from other decomposition methods (Sects. \ref{sec_basicdecomp} and \ref{sec_fullspecfit}), 
reveals the richest structural details relative to the input dust image (Figure \ref{fig:marks}). 
For instance, it most accurately reproduces the very long dust gap that extends across the north-south direction (the long black line in Figure \ref{fig:marks}) 
and the faint edge along the eastern wedge (cyan line in Figure \ref{fig:marks}). 
The extended clumps at the northern and southern tips of the eastern wedge, as well as the northern C-shaped structure previously highlighted in residual maps, 
are also recovered with minimal residuals. 
This suggests that these very narrow \ion{H}{I} components contribute a non-negligible fraction of the \ion{H}{I}-associated dust.

\subsection{Velocity distribution of dust-related \ion{H}{I} and CO}
\label{sec_velodist}

To investigate the dynamical correlation between the spectrally sharpened and positive-thresholded \ion{H}{I} ($\mathcal{H}_{\rm sharp}'$) and $^{12}$CO, we compare the velocity distributions of $\mathcal{H}$ (the original \ion{H}{I} cube), $\mathcal{H}_{\rm sharp}'$, and $^{12}$CO. 
For $\mathcal{H}$, the velocities are derived from its moment-1 map over the narrow component velocity range (-8 to 5 km s$^{-1}$), denoted as $V(\mathcal{H})$. For $^{12}$CO, we adopt the moment-1 map from Paper I, denoted as $V(^{12}\mathrm{CO})$. For $\mathcal{H}_{\rm sharp}'$, we adopt the moment-1 map weighted by $\alpha_{\beta=0.001}$ (from the regularized full-spectrum fitting), given by
\begin{equation}
    V(\mathcal{H}_{\rm sharp}') = \frac{ \sum_{-8<V_s<5} \alpha_{\beta=0.001;\ s}
    \mathcal{H}_{\rm sharp;\ s}' V_s }{ \sum_{-8<V_s<5} \alpha_{\beta=0.001;\ s}
    \mathcal{H}_{\rm sharp;\ s}'  }.
\end{equation}

Figure \ref{fig_velos} presents images of $V(\mathcal{H})$, $V(\mathcal{H}_{\rm sharp}')$, and $V(^{12}\mathrm{CO})$. 
The velocity distribution of $V(\mathcal{H})$ is strongly affected by broad emission components, 
resulting in an overall velocity range that is significantly narrower than those of $V(\mathcal{H}_{\rm sharp}')$ and $V(^{12}\mathrm{CO})$. 
In contrast, $V(\mathcal{H}_{\rm sharp}')$ and $V(^{12}\mathrm{CO})$ exhibit similar velocity ranges and 
show overall comparable spatial trends. 
Notably, the redshifted nature of Complex D (Figure \ref{fig:marks}) is clearly revealed in $V(\mathcal{H}_{\rm sharp}')$, 
with a spatial distribution closely matching that of $V(^{12}\mathrm{CO})$. 
Complexes D and AW (the southwestern quarter of the Polaris Flare; Figure \ref{fig:marks}) 
display striation patterns with alternating red- and blueshifted velocities 
when viewed in both $V(\mathcal{H}_{\rm sharp}')$ and $V(^{12}\mathrm{CO})$, 
highlighting the strong dynamical correlation among different complexes.

\begin{figure}
    \centering
    \includegraphics[width=0.9\linewidth]{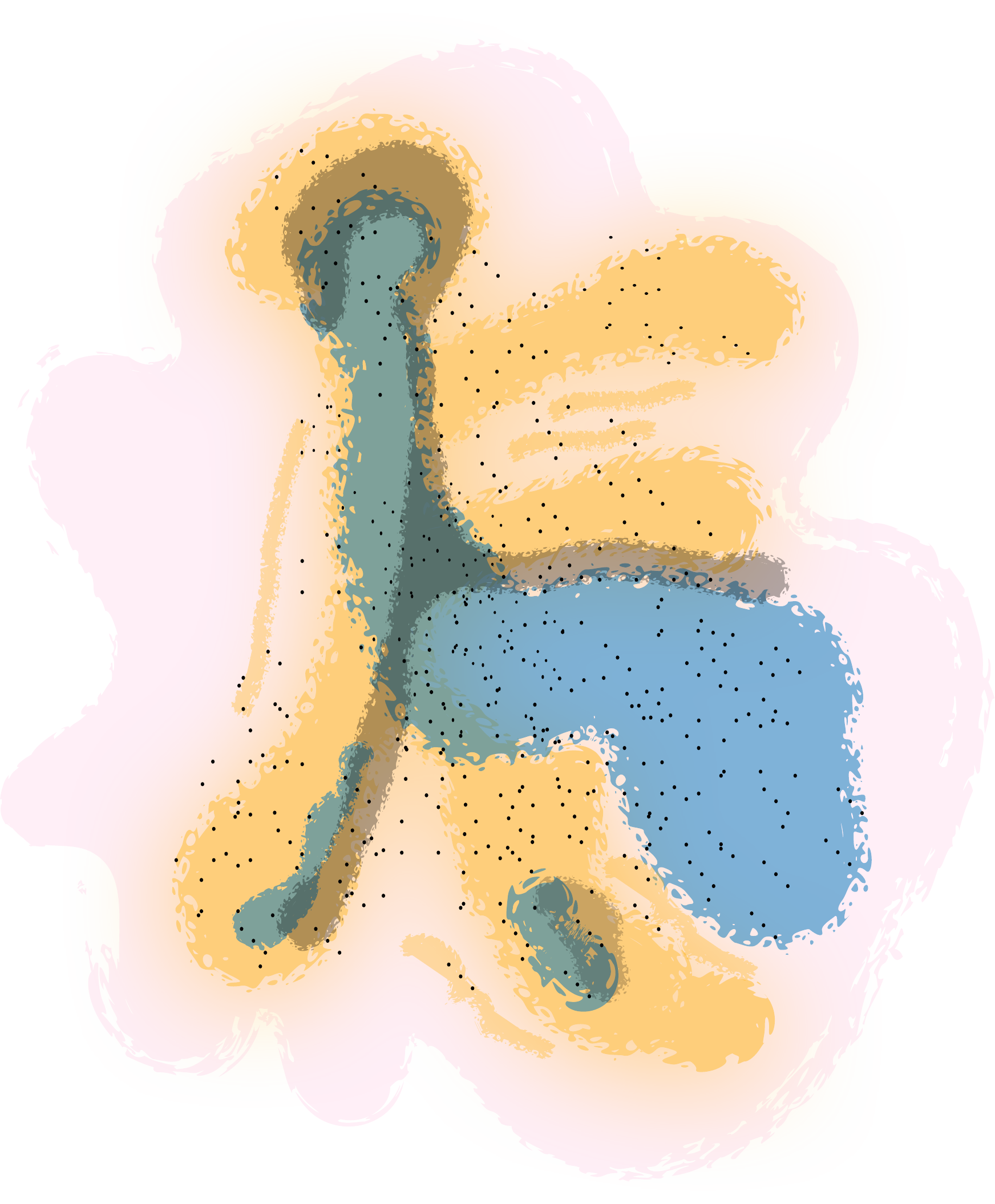}
\caption{Schematic sketch of the Polaris Flare, intended as a conceptual illustration. 
The extended pink region represents the diffuse broad \ion{H}{I} component (WNM). 
Orange regions show the narrow \ion{H}{I} component (CNM), including the very narrow feature interacting with the CO emission region. 
Blue regions indicate CO emission, while gray strips mark the CO- and \ion{H}{I}-dark zones at the atomic-to-molecular interfaces. 
Scattered black dots represent the distribution of dust.}
    \label{fig:sketch}
\end{figure}

On a global scale, the most prominent feature in these moment-1 maps is that the eastern wedge, the northern C-shaped structure, and Complex D together form a redshifted envelope enclosing the northern and eastern boundaries of the main CO emission region (Complex A). This configuration indicates that dust-related \ion{H}{I} and CO are dynamically associated, both spatially and spectrally, supporting the idea that the CNM, and possibly even colder components, are the major contributors to the \ion{H}{I}-associated dust. Furthermore, it suggests the presence of a compressed layer at the interface between the overall blueshifted CO emission region and the predominantly redshifted normal CNM.
Previous studies of nearby molecular clouds, such as Perseus, found no strong spatial correlation between CO and \ion{H}{I} \citep[e.g.,][]{2012ApJ...748...75L}, likely due to the more complex \ion{H}{I} emission at lower Galactic latitude and the lack of explicit separation of narrow CNM components. \ion{H}{I} narrow-line self-absorption (HINSA) studies in Taurus and other nearby clouds have independently shown that cold \ion{H}{I} is indeed mixed with molecular gas \citep[e.g.,][]{2005ApJ...622..938G,2010ApJ...724.1402K,2022A&A...658A.140L}, and analyses in more distant or massive clouds \citep[e.g.,][]{2024ApJ...973L..27S} also reveal excess atomic gas. However, these HINSA studies typically focus on single beams or limited regions and therefore cannot map the full spatial distribution of the excess atomic gas beyond the equilibrium expectation of $\sim$1~cm$^{-3}$ within molecular regions.
The Polaris Flare, located at high latitude with relatively simple line-of-sight structure, allows us to bridge these approaches by combining decomposition with dust residuals: we directly map the spatial morphology of the compressed interface, where excess cold \ion{H}{I} and CO-dark molecular gas coexist. The possible compression at the interface in Polaris likely leads to the formation of the very narrow (channel-width) \ion{H}{I} component and harbors CO-dark molecular gas, as traced by the dust residuals of the fitting.

\section{Discussion}\label{sec_diss}

\subsection{Zoom-in on residual dust at the \ion{H}{I}--CO interface}
The dust residuals reveal a distinct material component unaccounted for by either \ion{H}{I} or CO emission (Figure~\ref{fig_finalfit}). While the complex morphology along the eastern boundary (cyan line in Figure~\ref{fig:marks}) complicates a definitive interpretation, the main C-shaped feature provides an unambiguous example (Section~\ref{sec_detrendcross}). This well-defined residual structure is spatially nested between the integrated narrow \ion{H}{I} arc to the north and the northern perimeter of the molecular CO complex AW to the south (Figure~\ref{fig:mainC}).

This stratified configuration indicates that the residual dust tracks a transitional gas phase that is dark in both CO and 21~cm emission. This phase maps to CO-dark molecular gas within the atomic-to-molecular transition zone where H$_2$ has accumulated but CO remains photodissociated or sub-thermally excited. This localized accumulation highlights an interface layer sculpted by dynamical compression along the large-scale velocity gradient and by coherent multi-phase turbulence (Section~\ref{sec_velodist}). Quantitative decomposition shows that this interface component contributes $\ge$25\% of the local dust emission and accounts for 4--10\% of the global cloud dust mass (Table~\ref{tab:dustfractions}). These constraints provide direct observational evidence of a CO-dark molecular layer bridging the atomic and molecular distributions. This highlights the importance of multi-phase dynamical coherence, expanding upon previous single-phase frameworks \citep[e.g.,][]{1998A&A...336..697S,2025arXiv250220458L,2025NatAs...9.1366L}. To synthesize the spatial layout of these stratified gas phases and their joint dynamics, we present a conceptual schematic model in Section~\ref{sec:schematic}.

\subsection{Stepwise schematic map} 
\label{sec:schematic}

To synthesize the decomposition results, we present a schematic map of the Polaris Flare (Fig.~\ref{fig:sketch}), illustrating the stratified gas phases and their dynamical interplay. The broad \ion{H}{I} component traces the WNM, providing a diffuse, low-dust baseline, whereas the narrow \ion{H}{I} component traces the CNM, contributing dust in extended atomic structures and in regions adjacent to molecular gas. At the H\,\textsc{i}--H$_2$ interface, a distinct CNM layer closely surrounds the CO-bright region, marking the transition to molecular gas. Beyond the CO-bright zone, a CO-dark H$_2$ layer contributes roughly 10--25\% of the dust at the interface, while molecular gas traced by $^{12}$CO, $^{13}$CO, and C$^{18}$O contains the bulk of the remaining dust. This sequence illustrates a progression from diffuse WNM, to normal CNM, to interface CNM, to CO-dark H$_2$, and finally to CO-bright molecular gas, demonstrating how dust traces successive phases and complements the velocity information discussed above.  

Superimposed on this static hierarchy is a coherent dynamical configuration revealed by the velocity distributions. The eastern wedge, northern C-shaped structure, and Complex D form a redshifted envelope surrounding the main CO emission (Sect. \ref{sec_velodist}), indicating dynamical interaction between CNM and CO gas. This interaction produces a compressed interface layer, likely giving rise to the very narrow (channel-width) \ion{H}{I} component and promoting CO-dark H$_2$ accumulation, as independently traced by the dust residuals. The observed increase in dust content in progressively colder atomic gas suggests that H$_2$ formation is facilitated by dust shielding, while dust growth itself may be coupled to molecule formation. Although the schematic is specific to Polaris, the conceptual framework of stepwise decomposition and multi-phase dynamical interaction is broadly applicable to other interstellar clouds, with the relative contributions of each phase varying according to evolutionary stage, ambient radiation, and stellar feedback. Further validation will benefit from extensive CO surveys at high Galactic latitudes, complemented by \ion{H}{I} observations \citep[e.g.,][]{2013PASA...30....3D,2018IMMag..19..112L,2024yCatp073006701Z} at spatial and spectral resolutions  matched to dust and CO data to disentangle distinct neutral components. This discussion naturally leads to an assessment of the limitations and assumptions of the overall decomposition method.

\subsection{Caveats and limitations}

While our decomposition and schematic provide a useful conceptual framework, several simplifying assumptions and observational limitations affect the quantitative interpretation across the entire study. The analysis assumes a globally constant linear response between gas and dust, which may not always hold, especially considering the evolution of dust properties both cloud-to-cloud and within individual clouds \citep[e.g.,][]{2010A&A...518L..74R,2015ApJ...811..118R,10.1093/mnras/stx118,2018ApJ...862..131M}.For instance, the brightest CO emission regions, affected by optical thickness, can contribute to dust residuals, and the linear relation between CO and dust in the \ion{H}{I} components is likely more complex than represented by the simplified two-phase (WNM and CNM) or three-phase (including the very narrow CNM) model. Achieving a more accurate decomposition requires \ion{H}{I} observations with higher spatial and spectral resolution to disentangle the WNM and CNM distributions beyond the approximation of integrating over broad velocity ranges, alongside dedicated, potentially non-linear analyses rather than global linear fitting.  

The spatial offset observed between \ion{H}{I} and CO emission at their interface strongly suggests the presence of CO-dark molecular gas. However, the limited sensitivity of CO, compared to the typically high signal-to-noise \ion{H}{I} data, complicates definitive confirmation. In Paper I, we show that the 4~K isosurface encloses the main CO emission body in the PPV cube, while the region between the 2~K and 4~K isosurface is very narrow, indicating a sharp boundary. Based on this evidence, the dust residuals in these interface regions are most reasonably interpreted as tracing CO-dark molecular gas, though future observations with improved sensitivity, higher resolution, and more sophisticated non-linear analyses will be essential to confirm this assignment definitively and to refine the overall decomposition framework introduced in this work.

\section{Summary}\label{sec_summary}

By adopting dedicated linear decomposition techniques, including different linear fitting combinations, full-spectrum fittings, and a regularization approach, we reconstruct the Planck dust map based on CO emission from the PMO Polaris CO survey (PPCOS) and the EBHIS \ion{H}{I} emission in the Polaris Flare. A considerable fraction of residual dust remains that cannot be fully accounted for by either \ion{H}{I} or CO. 
The fitting results are verified to be robust despite the relatively lower spatial resolution of \ion{H}{I}.
We quantify the contributions of various gas phases to dust emission and examine their kinematic interplay. The main findings are:  

\begin{itemize}
    \item CO-associated dust contributes 20--40\% of the total dust, whereas the WNM shows negligible dust content. The derived $X_{\rm ^{12}CO}$ value is $1.2 \times 10^{22}$~cm$^{-2}$~K$^{-1}$~km$^{-1}$~s, consistent with literature estimates.
    
    \item For the \ion{H}{I} component, dust content increases along the evolutionary sequence of \ion{H}{I} components, suggesting progressive molecule formation and associated dust growth. \ion{H}{I}-associated dust is primarily concentrated in the CNM, traced by the narrow \ion{H}{I} component, and in even colder neutral gas indicated by the very narrow \ion{H}{I} component. 
    
   \item Residual dust is primarily concentrated along the eastern edge of the Polaris Flare, accounting for 4--10\% of the total dust mass. Zoom-in views of the residual map, particularly around the main C-shaped feature, reveal a narrow morphology located at the interface between \ion{H}{I} and CO emission. This distribution is consistent with the presence of CO-dark molecular gas at atomic--molecular (\ion{H}{I}--CO) interfaces.
    
    \item \ion{H}{I}-associated dust is assigned to individual velocity channels through regularized full-spectrum linear fitting. The spectrally sharpened \ion{H}{I} cube, weighted by the fitted parameters ($\alpha_{\beta=0.001}$), reproduces well the global \ion{H}{I} velocity gradient observed in CO, capturing both the spatial trend and amplitude. This confirms that a significant fraction of \ion{H}{I}-associated dust resides in the very narrow \ion{H}{I} components. The overall redshifted \ion{H}{I} enveloping the blueshifted CO emission further suggests dynamic interaction between the CO-emitting gas and the surrounding CNM.
    
   \item Motivated by the observational and decomposition results above, we present a stepwise schematic map synthesizing our findings. The map depicts the stratified gas phases from diffuse WNM, through normal and interface CNM, to CO-bright molecular gas, highlighting the CO-dark H$_2$ at atomic--molecular interfaces. It provides a conceptual framework for visualizing how dust traces successive phases and how multi-phase dynamics influence cloud structure.
\end{itemize}

Our analysis provides a component-resolved view of dust associated with CO, \ion{H}{I}, and CO-dark molecular gas in the Polaris Flare, revealing the coupling between atomic and molecular gas. This coupling is likely common but warrants further examination in broader environments. The multi-technique linear decomposition, as a promising framework for analyzing large-area CO, \ion{H}{I}, and dust distributions, offers a practical approach to examine the structure and dust content of multi-component ISM.

\begin{acknowledgement}
X.L. acknowledges the support of the Strategic Priority Research Program of the Chinese Academy of Sciences  under Grant No. XDB0800303.
T.Z. acknowledges the Leading Innovation and Entrepreneurship Team of Zhejiang Province of China (Grant No. 2023R01008).
Z.H was supported by NSFC through grants 12573023,12303024, and the Natural Science Foundation of Sichuan Province (2024NSFSC0453). 
This research was carried out in part at the Jet Propulsion Laboratory, which is operated by the California Institute of Technology under a contract with the National Aeronautics and Space Administration (80NM0018D0004).
We thank the staff of the Delingha Observatory for their continuous efforts in carrying out the observations of the PMO Polaris CO survey.
\end{acknowledgement}

\bibliography{PMOPolarisXco}
\bibliographystyle{aa} 

\begin{appendix}
\section{Bilinear interpolation of irregular data}\label{sec_bilinear}
Bilinear interpolation is a widely used method for reprojecting images.  
It provides a simple and efficient way to obtain smoothly sampled data with minimal loss of spatial resolution, provided that the input data are regularly gridded.  
However, when the data are irregularly gridded, such as the HEALPix format used in EBHIS or even more irregularly sampled datasets, preprocessing is required to regularize the input before bilinear interpolation can be applied effectively.  

\subsection{Bilinear interpolation of regular data}
We begin by considering the case of bilinear interpolation on a regular grid.  
Assume four points, denoted $p_1$, $p_2$, $p_3$, and $p_4$,  
that form a square. Their coordinates are given by  
\begin{equation}
\begin{aligned}
x_{p_1} &= x_{p_3} = x_l, & \quad x_{p_2} &= x_{p_4} = x_r, \\
y_{p_1} &= y_{p_2} = y_b, & \quad y_{p_3} &= y_{p_4} = y_u.
\end{aligned}
\label{eq:coords}
\end{equation}
Here $p_1=(x_l,y_b)$ is the lower--left corner,  
$p_2=(x_r,y_b)$ the lower--right corner,  
$p_3=(x_l,y_u)$ the upper--left corner,  
and $p_4=(x_r,y_u)$ the upper--right corner.  

Denote $I_i$ as the value at $p_i$.  
For a given resampling point located within that square, with coordinates $(x_0, y_0)$,
the bilinear interpolation yields
\begin{equation}
    I_0 = \sum_{i=1}^{4} w_i I_i,
\end{equation}
with
\begin{equation}
\begin{aligned}
    w_1 &= (1-\alpha)(1-\beta), & w_2 &= \alpha(1-\beta),\\
    w_3 &= (1-\alpha)\beta,     & w_4 &= \alpha \beta,
\end{aligned} \label{eq_wform}
\end{equation}
and
\begin{equation}
    \alpha=\frac{x_0-x_l}{x_r-x_l}, \qquad 
    \beta=\frac{y_0-y_b}{y_u-y_b}. \label{eq_alphabeta}
\end{equation}
By construction, the weights are normalized:
\begin{equation}
    \sum_{i=1}^4 w_i = 1.
\end{equation}
Similarly, the coordinate of $p_0$ satisfies
\begin{equation}
    p_0 = \sum_{i=1}^4 w_i p_i. \label{eq_interpcoord}
\end{equation}

\begin{figure}
    \centering
    \includegraphics[width=0.9\linewidth]{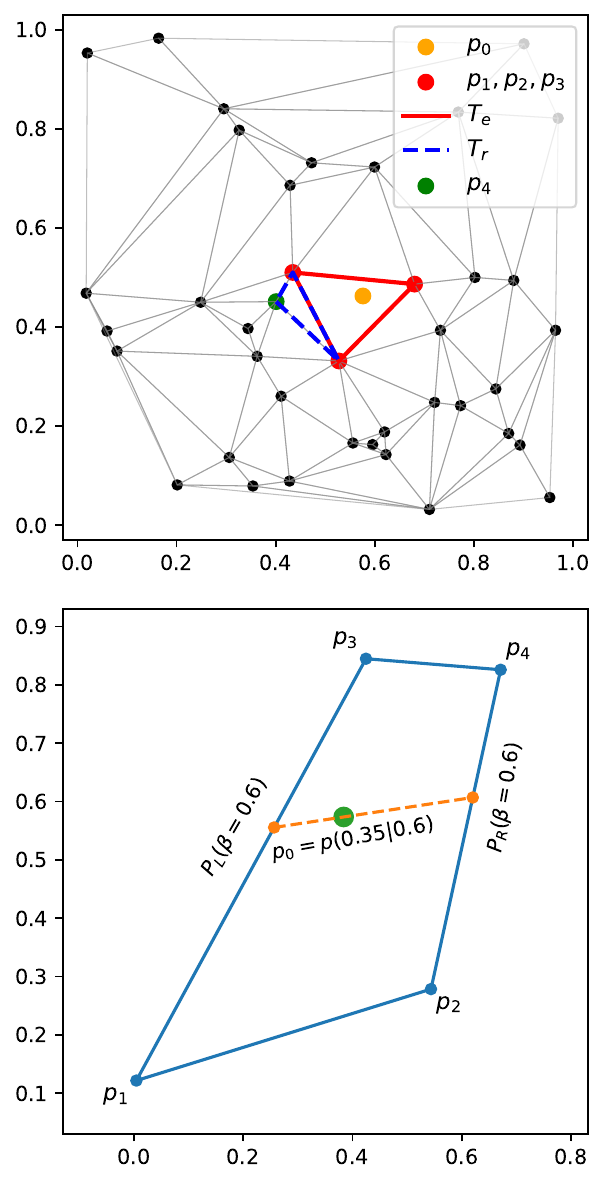}
    \caption{Upper: Illustration of finding the enclosing triangle for a given point $p_0$ and its regular neighbor. Black dots represent input samples, which are randomly scattered here, and gray lines show the Delaunay triangulation. 
     Lower: Schematic showing how the parameter set ($\alpha$, $\beta$) uniquely determine the location of a target point within a convex quadrilateral.
     \label{fig_general_bilinear}
}

    \label{fig:bilinear}
\end{figure}

\begin{figure*}
    \centering
    \includegraphics[width=0.99\linewidth]{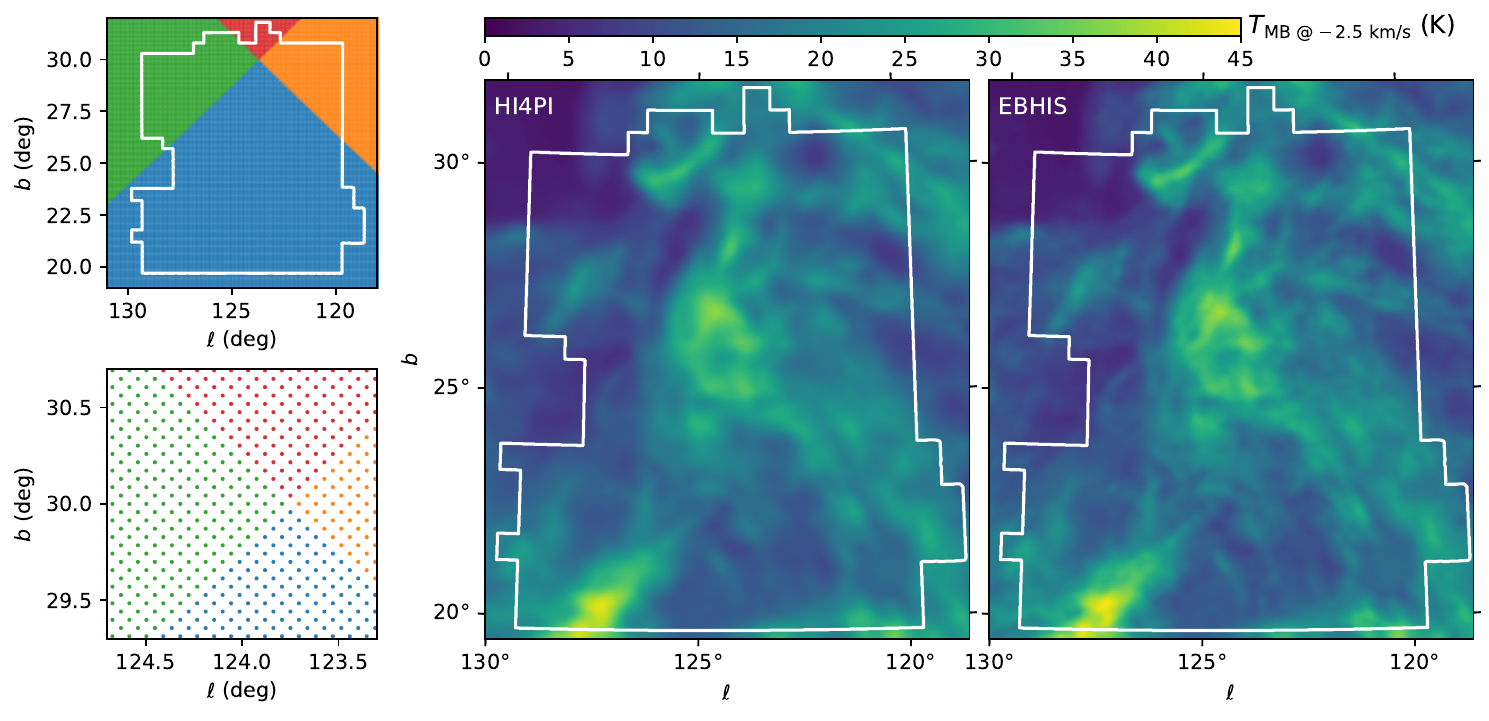}
    \caption{
    Left Upper: Sample points of the EBHIS HEALPix data toward the Polaris region. Different colors represent different regions defined by the coarse Nside=4 grid.  
    Left Lower: Zoom-in of the left upper panel.  
    Middle: Channel map at $-2.5~\mathrm{km~s^{-1}}$ interpolated from the HI4PI data.  
    Right: Same as the middle panel, but for the HI spectra interpolated from the EBHIS data.  
    In the left upper, middle, and right panels, the white contours indicate the sky coverage of the PMO Polaris CO survey. \label{fig_HI4PI_EBHIS}
}

\end{figure*}

\subsection{Triangulation and regular neighbors}
For regularly sampled input data, it is straightforward to identify a square (or rectangular) cell that encloses any given point.  
However, for irregularly sampled data, this becomes more challenging.  
A common solution is to apply Delaunay triangulation\footnote{For example, implemented in \texttt{scipy.spatial.Delaunay}} \citep{Ito2015} to the input points (upper panel of Figure \ref{fig_general_bilinear}).
Once the triangulation is constructed, any given point can be uniquely located within one of the triangles defined by the triangulation.  
Interpolation using the values at the three vertices of the enclosing triangle is a possible approach, but it relies on only three points and does not naturally reduce to bilinear interpolation in the limit of regular input data.  
For instance, consider a point $p_0$ located at the center of a square defined by four points, $p_1$–$p_4$, as in Eq.~\ref{eq:coords}.  
In the case of bilinear interpolation on a regular grid, all four points would contribute equally to the interpolated value.  
However, using a triangulation-based approach, only three of these points are assigned as vertices of the enclosing triangle, which can lead to unequal weighting and deviation from the expected bilinear result.

Then, the task becomes finding a fourth point ($p_4$) such that, together with the three vertices of the enclosing triangle ($T_e$, defined by $p_1$, $p_2$, and $p_3$), they form a quadrilateral that is as regular as possible.  
Naturally, this fourth point can be chosen as the additional vertex of one of the triangles neighboring the enclosing triangle.  
We refer to this neighboring triangle as the \emph{regular neighbor}.  
For data points confined to a 2D plane, there are at most three neighboring triangles ($\mathcal{N}$).  
A simple criterion to identify the regular neighbor is to select the vertex that maximizes the angle opposite the shared edge, excluding the vertices of $T_e$:
\begin{equation}
p_4 = {\rm argmax}_{p} \{\ \angle p \,|\, p \in \mathcal{N},\ p \notin T_e \ \}.
\end{equation}
Denote the neighbor containing $p_4$ as the regular neighbor ($T_r$).  
Without loss of generality, assume $p_1 \notin T_r$.  
Typically, ($p_1$, $p_2$, $p_4$, $p_3$) form a convex quadrilateral, provided that the input data are not too irregularly sampled. 
Then $p_1$, $p_2$, $p_3$, and $p_4$ can be treated as the vertices of the regular quadrilateral (upper panel of Figure \ref{fig_general_bilinear}), ordered similarly as in Eq.~\ref{eq:coords}.

\subsection{Generalized bilinear triangulation}
The generalized version of bilinear triangulation consists of finding the weights $w_i$, 
in the form of Eq.~\ref{eq_wform}, that satisfy the coordinate reconstruction condition, 
Eq.~\ref{eq_interpcoord}.
Consider a convex quadrilateral with vertices $p_1, p_2, p_3, p_4$, and a point $p_0$ inside it.  
The bilinear interpolation formula reads
\begin{equation}
p_0 = (1-\alpha)(1-\beta)p_1 + \alpha(1-\beta)p_2 + (1-\alpha)\beta p_3 + \alpha \beta p_4.
\end{equation}
Define points on the left and right edges as
\begin{equation}
p_L(\beta) = (1-\beta)p_1 + \beta p_3, \qquad
p_R(\beta) = (1-\beta)p_2 + \beta p_4.
\end{equation}
Any point on the segment connecting $p_L(\beta)$ and $p_R(\beta)$ can be written as
\begin{equation}
p(\alpha|\beta) = (1-\alpha)p_L(\beta) + \alpha p_R(\beta), \quad 0 \le \alpha \le 1.
\end{equation}
To satisfy the bilinear formula, $p_0$ must lie on this line segment:
\begin{equation}
p_0 - p_L(\beta) = \alpha \bigl(p_R(\beta)-p_L(\beta)\bigr).
\end{equation}
For a convex quadrilateral, $p_0$ moves continuously from $p_1$ to $p_3$ along $p_L(\beta)$ as $\beta$ increases from 0 to 1.  
Therefore, there exists a \emph{unique} $\beta \in (0,1)$ such that $p_0$ lies on the corresponding segment $[p_L(\beta), p_R(\beta)]$.  
Once $\beta$ is fixed, $\alpha$ is uniquely determined by projecting $p_0 - p_L(\beta)$ along $p_R(\beta)-p_L(\beta)$:
\begin{equation}
\alpha = \frac{(p_0 - p_L(\beta)) \cdot (p_R(\beta)-p_L(\beta))}{|p_R(\beta)-p_L(\beta)|^2}.
\end{equation}
This proves that the pair $(\alpha,\beta)$ is unique for any point $p_0$ inside a convex quadrilateral (see lower panel of Figure \ref{fig_general_bilinear}).

Explicitly solving for $(\alpha, \beta)$ is not straightforward for an arbitrary convex quadrilateral.  
In practice, we determine them iteratively using a Newton-Raphson solver.  
Starting from an initial guess $(\alpha_0, \beta_0) = (0.5, 0.5)$, the solver updates the parametric coordinates according to
\begin{equation}
F(\alpha, \beta) = 
\begin{pmatrix}
x(\alpha,\beta) - x_0 \\
y(\alpha,\beta) - y_0
\end{pmatrix}, \quad
(\alpha,\beta) \leftarrow (\alpha,\beta) - J^{-1} F(\alpha,\beta),
\end{equation}
where $J$ is the Jacobian of the mapping from $(\alpha,\beta)$ to the physical coordinates.  
The iteration proceeds until $F$ is smaller than a tolerance, yielding the bilinear weights $w_i$ via Eq.~\ref{eq_wform}.
If the four vertices form a square, this method reduces back to the normal bilinear interpolation with $(\alpha, \beta)$ described by Eq.~\ref{eq_alphabeta}.

\subsection{Bilinear interpolation of EBHIS data}

The EBHIS data \citep{2016A&A...585A..41W}, when incorporated into the HI4PI product \citep{2016A&A...594A.116H}, have been smoothed to match the resolution of the southern-sky GASS survey \citep{2009ApJS..181..398M}. Here, we instead use the EBHIS data directly to retain the higher native spatial resolution.  
The EBHIS survey provides data in both equatorial and Galactic coordinates.  
Because distortions are severe near the pole, we adopt the Galactic-coordinate data, which are stored in the HEALPix format.  
The released maps use \texttt{Nside} = 1024, corresponding to an effective pixel size of $\simeq 3.4'$ (pixel area $\simeq 11.8~\mathrm{arcmin}^2$).  
To save storage, the full sky is partitioned into 192 subregions (the \texttt{Nside} = 4 base pixels), with each subregion distributed as a separate file. The left panels Figure~\ref{fig_HI4PI_EBHIS} show the sampled points in the region covering the Polaris Flare. The data appear regularly gridded, but in a rotated orientation.
We apply bilinear interpolation to extract a subcube covering the Polaris region from both the EBHIS HEALPix data and the HI4PI dataset. In the following, we refer to these extracted subcubes simply as the EBHIS data and HI4PI data, respectively, without ambiguity.

The middle and right panels of Figure~\ref{fig_HI4PI_EBHIS} show channel maps at $-1.2~\mathrm{km~s^{-1}}$, corresponding to the brightest channel, for the HI4PI and EBHIS data, respectively.  
The EBHIS map appears sharper than the HI4PI map, although the difference is modest.  
To quantify the relative sharpness of the two images, we employ the variance of the Laplacian operator as a sharpness metric.  
The Laplacian emphasizes spatial intensity gradients, and its variance provides a scalar estimate of the contrast of structural edges in the image, with larger values corresponding to sharper, better-resolved structures \citep[e.g.,][]{Pech2000}.  
For a given image $I$, the sharpness is defined as
\begin{equation}
S(I) = \mathrm{Var}\!\left( I \ast L \right),
\end{equation}
where $L$ denotes the $3\times 3$ discrete Laplacian kernel and $\ast$ represents convolution.  
To compare two images, $I_1$ and $I_2$, we define the normalized sharpness ratio
\begin{equation}
R = \frac{S(I_1)}{S(I_2)} \; \left( \frac{\sum I_2}{\sum I_1} \right)^{2},
\end{equation}
where the flux normalization term accounts for minor differences in total intensity resulting from the non-strict flux conservation of interpolation.  
By construction, $R > 1$ indicates that $I_1$ is sharper than $I_2$, while $R < 1$ implies the opposite.  
The computed sharpness ratio between the EBHIS and HI4PI data is $R \sim 1.2$, indicating that the EBHIS data resolves finer structures, albeit with slightly higher noise, as expected. This is because the HI4PI data have been smoothed, whereas the bilinear interpolation preserves the higher native resolution of the EBHIS data.

\section{Matrix formulation of the fitting} \label{app_matrix}
\begin{figure}
    \centering
    \includegraphics[width=0.99\linewidth]{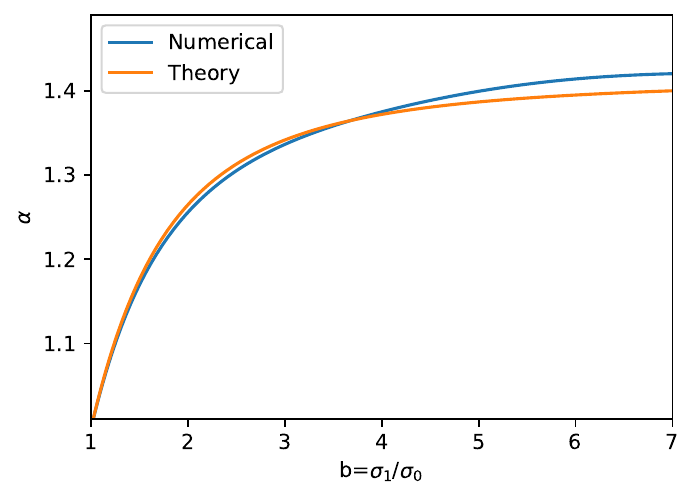}
    \caption{Comparison of the $\alpha$ values from numerical calculation (blue) and theoretical prediction (orange) as a function of the beam ratio in the 1D case.}
    \label{fig:resolutioneffect}
\end{figure}
For self-consistency, we briefly present the matrix form of the least-squares fitting. 
The standard derivation can be found in textbooks \citep[e.g.,][]{strang09}. 

Let $\mathcal{D}$ be the objective map (serving as a vector of $N$ pixels) and $I_1, \dots, I_n$ the fitting maps, 
with $I_{n+1} = 1$ a constant map. Define the design matrix
\begin{equation}
    \mathbf{A} = 
    \begin{bmatrix}
    I_1 & I_2 & \cdots & I_n & I_{n+1}
    \end{bmatrix}_{N \times (n+1)},
\end{equation}
and the coefficients
\begin{equation}
    \boldsymbol{\alpha} = 
    \begin{bmatrix}
    \alpha_1 \\ \alpha_2 \\ \vdots \\ \alpha_n \\ \alpha_c
    \end{bmatrix}_{(n+1) \times 1}.
\end{equation}
The correlation (Gram) matrix is
\begin{equation}
    \mathbf{C} \equiv \mathbf{A}^{\rm T} \mathbf{A}.
\end{equation}

The least-squares problem is then
\begin{equation}
    \boldsymbol{\alpha} = \underset{\boldsymbol{\alpha}}{\rm argmin} \; 
    \left\| \mathcal{D} - \mathbf{A} \boldsymbol{\alpha} \right\|^2,
\end{equation}
with the solution
\begin{equation}
    \boldsymbol{\alpha} = \mathbf{C}^{-1} \mathbf{A}^{\rm T} \mathcal{D}.\label{eq_alphasolv}
\end{equation}
Here, $\mathbf{A}^{\rm T} \mathcal{D}$ gives the correlation between 
$\mathcal{D}$ and the fitting maps.
The fitted map is
\begin{equation}
    \mathcal{D}_{\rm fit} = \mathbf{A} \boldsymbol{\alpha}.
\end{equation}

\subsection{Mean-subtraction trick} 
Subtract the mean from each map,
\begin{equation}
    \bar{\mathcal{D}} = \mathcal{D} - \langle \mathcal{D} \rangle, \quad
    \bar{I}_s = I_s - \langle I_s \rangle, \quad s=1,\dots,n,
\end{equation}
denoting mean-subtracted maps with a bar. After this, the constant map is no longer needed. Define the mean-subtracted correlation matrix
\begin{equation}
    \bar{\mathbf{C}} \equiv \bar{\mathbf{A}}^{\rm T} \bar{\mathbf{A}}, \quad 
    \bar{\mathbf{A}} = [\bar{I}_1, \dots, \bar{I}_n].
\end{equation}

The matrix solution then reduces to
\begin{equation}
    \boldsymbol{\alpha} = \bar{\mathbf{C}}^{-1} \bar{\mathbf{A}}^{\rm T} \bar{\mathcal{D}},
\end{equation}
with the fitted map
\begin{equation}
    \bar{\mathcal{D}}_{\rm fit} = \sum_{s=1}^{n} \alpha_s \bar{I}_s,
\end{equation}
and the full fitted map including the mean is
\begin{equation}
    \mathcal{D}_{\rm fit} = \bar{\mathcal{D}}_{\rm fit} + \langle \mathcal{D} \rangle.
\end{equation}

\subsection{Influence of resolution} \label{sec_theoryaboutbeameffect}
To show the effect of unmatched resolution, we consider a simple case with a single fitting vector $\mathbf{e}$.  
The least-squares solution is
\begin{equation}
    \alpha = \frac{\mathbf{e}^T \mathcal{D}}{\mathbf{e}^T \mathbf{e}}. \label{verysimpletest}
\end{equation}
For simplicity, let
\begin{equation}
    \mathcal{D} \equiv \mathbf{e} = (\xi * G_{\sigma_0})(x),
\end{equation}
where $\xi(x)$ is a white-noise process and $G_{\sigma_0}(x)$ is a Gaussian kernel with standard deviation $\sigma_0$.  
In this case, the solution yields $\alpha = 1$.

Now, if we smooth $\mathbf{e}$ with a Gaussian kernel of width $\sigma_1$, the resulting vector is
\begin{equation}
    \mathbf{e}_{\rm sm} = (\xi * G_{\sigma_1})(x),
\end{equation}
and inserting $\mathbf{e}_{\rm sm}$ into Eq.~\ref{verysimpletest} gives
\begin{equation}
    \alpha = \frac{\int G_{\sigma_0}(x) G_{\sigma_1}(x) \, dx}{\int G_{\sigma_1}(x)^2 \, dx} 
           = \sqrt{\frac{2 \sigma_1^2}{\sigma_0^2 + \sigma_1^2}}
           = \sqrt{\frac{2 b^2}{1 + b^2}},
\end{equation}
where $b \equiv \sigma_1 / \sigma_0$ is the ratio of the smoothing kernels.  
As a demonstration, we generate a white-noise sequence and apply Gaussian smoothing with different widths. The resulting values of $\alpha$ as a function of $b$ are shown in Figure~\ref{fig:resolutioneffect}, confirming the analytical expression above.

For a 2-dimensional map, the corresponding expression becomes
\begin{equation}
    \alpha = \frac{2 b^2}{1 + b^2}.
\end{equation}
Thus, in 2D, $\alpha$ is at most doubled when fitting with an image of lower resolution.

\subsection{Regularization of linear fitting}\label{sec_aboutregularization}
When the correlation matrix $\mathbf{C}$ is ill-conditioned, i.e., some eigenvalues are very small compared to the largest eigenvalue $\Lambda_{\rm max}(\mathbf{C})$ due to highly correlated maps, the solution $\boldsymbol{\alpha}$ from Eq.~\ref{eq_alphasolv} can exhibit large coefficients and spurious fluctuations along directions associated with these small eigenvalues.  

To stabilize the solution, we introduce Tikhonov (ridge) regularization \citep{tikhonov1977solutions}, which adds a positive term $\lambda$ to the diagonal of $\mathbf{C}$:
\begin{equation}
    \mathbf{C}_\lambda = \mathbf{C} + \lambda \mathbf{I}, \label{eq_Clambda}
\end{equation}
and modifies Eq.~\ref{eq_alphasolv} to
\begin{equation}
    \boldsymbol{\alpha}_\lambda = \mathbf{C}_\lambda^{-1} \mathbf{A}^{\rm T} \mathcal{D}, \label{eq_alphalambda}
\end{equation}
where $\lambda > 0$ is the regularization parameter and $\mathbf{I}$ is the identity matrix.

This procedure effectively increases the eigenvalues of $\mathbf{C}$, suppressing the contribution of modes associated with small eigenvalues.  
When $\lambda = 0$, the solution reduces to standard linear fitting without regularization, whereas larger values of $\lambda$ push the coefficients $\boldsymbol{\alpha}$ closer to zero overall.  
As a result, both the magnitude of the coefficients and their sensitivity to noise are reduced, yielding a more stable and physically meaningful solution.  
There is no universal rule for choosing $\lambda$, but a practical choice is
\begin{equation}
    \lambda = \beta \, \Lambda_{\rm max}(\mathbf{C}),
\end{equation}
where $\beta \sim 10^{-3}$ is a dimensionless parameter calibrated to the overall scale of $\mathbf{C}$.

In general, regularized fitting produces a stabilized solution ($\alpha_s$) that is smoother and less sensitive to ill-conditioning.  
For the present work, we adopt this method to obtain a robust decomposition of the dust map into its atomic and molecular components.

\end{appendix}

\end{document}